\documentclass[twocolumn,letterpaper]{IEEEconf}
\usepackage[flushleft]{threeparttable}
\usepackage{cite}
\usepackage{times}
\usepackage{amsmath}
\usepackage{psfrag}
\usepackage{bm}
\usepackage{dblfloatfix}
\usepackage{mathtools}
\usepackage{amssymb}
\usepackage{subfigure}
\usepackage{verbatim}
\usepackage{graphicx}
\usepackage[thinlines,thiklines]{easybmat}
\usepackage{latexsym}
\usepackage[dvipsnames]{xcolor}
\usepackage{cite}
\usepackage{stmaryrd}
\usepackage{cuted}
\usepackage{lipsum}
\usepackage{float}
\usepackage{multirow}
\usepackage{textcomp}
\usepackage[linesnumbered,ruled,vlined]{algorithm2e}
\usepackage{bookmark}

\newsavebox{\twosubbox}

\newtheorem{lem}{Lemma}

\newtheorem{defn}{Definition}

\def\s{\mathop{\rm s}\nolimits}
\def\c{\mathop{\rm c}\nolimits}
\newcommand{\bs}[1]{\ensuremath{{\boldsymbol{#1}}}}

\def\diag{\mathop{\rm diag}\nolimits}

\def\diag{\mathop{\rm diag}\nolimits}

\makeatletter
\def\ALG@special@indent{%
	\ifdim\ALG@thistlm=0pt\relax
	\hskip-\leftmargin
	\else
	\hskip\ALG@thistlm
	\fi
}
\newcommand{\Notations}[1]{\item[]\noindent\ALG@special@indent \textbf{Notations:}\ #1}

\makeatletter
\newcommand{\multiline}[1]{%
	\begin{tabularx}{\dimexpr\linewidth-\ALG@thistlm}[t]{@{}X@{}}
		#1
	\end{tabularx}
}
\makeatother

\DeclareFontFamily{OMX}{yhex}{}
\DeclareFontShape{OMX}{yhex}{m}{n}{<->yhcmex10}{}
\DeclareSymbolFont{yhlargesymbols}{OMX}{yhex}{m}{n}
\DeclareMathAccent{\wideparen}{\mathord}{yhlargesymbols}{"F3}

\title{\bf \Large Learning-Based Safe Motion Control of Vehicle Ski-Stunt Maneuvers~\thanks{This work was partially supported by the US National Science Foundation under award CNS-1932370.}}

\author{Feng Han and Jingang Yi~\thanks{F. Han and J. Yi are with the Department of Mechanical and Aerospace Engineering, Rutgers University, Piscataway NJ 08854 USA (email: fh233@scarletmail.rutgers.edu, jgyi@rutgers.edu).}}

\begin{document}
\maketitle

\begin{abstract}
This paper presents a safety guaranteed control method for an autonomous vehicle ski-stunt maneuver, that is,  a vehicle moving with two one-side wheels. To capture the vehicle dynamics precisely, a Gaussian process model is used as additional correction to the nominal model that is obtained from physical principles. We construct a probabilistic control barrier function (CBF) to guarantee the planar motion safety. The CBF and the balance equilibrium manifold are enforced as the constraints into a safety critical control form. Under the proposed control method, the vehicle avoids the obstacle collision and safely maintain the balance for autonomous ski-stunt maneuvers. We conduct numerical simulation validation to demonstrate the control design. Preliminary experiment results are also presented to confirm the learning-based motion control using a scaled RC truck for autonomous ski-stunt maneuvers.
\end{abstract}


\section{Introduction}
\label{Sec_Introduction}

Ski-stunt maneuver is a vehicle driving technique in which only two one-side wheels move on the ground and other two wheels are in the air. Ski-stunt maneuver is one of the unstable, agile maneuvers~\cite{YiCST2011} and vehicle motion may risk rolling over completely. The maneuver is usually performed by professional racing car drivers~\cite{howell2014monster} and the first ski-stunt maneuver was performed at the 1964 World Fair in Denmark. Multiple different ways are used to initialize a ski-stunt maneuver and a relatively safe and tractable method is to drive the vehicle on a ramp to lift one side and then maintain the tilted chassis after leaving the ramp. For vehicles with high center of gravity (e.g., the sport utility vehicles and trucks), drivers might initiate and perform a ski-stunt maneuver when turning sharply at certain high speed.

The vehicle motion under ski-stunt maneuvers is highly related to rollover control since it experience a large roll motion and motion initialization is in the same way~\cite{Ralf2000Rollover, Rollover2013Phanomchoeng}. Rollovers happen when the vehicles turn sharply (for example exiting a highway from a circular ramp) or when the vehicle hit on a small obstacle at high speed, that is, untripped and tripped rollovers~\cite{Rollover2013Phanomchoeng}. Comparing with the rollover accident, ski-stunt maneuver motion is in a balanced, safely controlled fashion. Therefore, study of the autonomous ski-stunt maneuver is of great importance for vehicle safety operation. Moreover, autonomous skit-stunt maneuvers can also be used as an active safety features for the next-generation of zero-accident design, such as under emergency situation for obstacle avoidance maneuver~\cite{Springfeldt1996Rollover, matolcsy2007severity, Steering2012Imine}.

When conducting a ski-stunt maneuver, the vehicle motion is underactuated with three degrees of freedom (DOFs), that is, roll motion as well as planar motion, but with two control inputs, that is, steering and velocity actuation. To prevent the vehicle from rollover or collision, safety guaranteed design must be considered. Safety critical control design by the control barrier function (CBF) method is an effective approach for autonomous robots and vehicle~\cite{Arab2021Acc, Seo2022Vehicle} and other balance robots~\cite{Ames2019Barrier,Khan2021Safety}. Safety critical control does not explicitly design any trajectory tracking inputs and instead, CBF is used as dynamic constraint to directly update the nominal control as a safety guaranteed certification. Therefore, CBF-based control can be applied in real time for obstacle avoidance design. Although the planar motion control cannot necessarily guarantee the stable roll motion due to the underactuated property, further attention is needed to maintain the roll balance motion.

When running on two one-side wheels during ski-stunt maneuver, the vehicle displays a single-track characteristic, similar to motorcycle or bicycle dynamics. Motion and balance control of autonomous motorcycles and bicycles have been reported in the past decades~\cite{GetzPhD,Lee2002,YiICRA2006,WangICRA2017,WangTASE2019}. Steering and velocity control are among the most effective actuation for autonomous bicycles. A balance equilibrium manifold (BEM) concept is used to capture the trajectory tracking and balance control simultaneously. All of the above-mentioned motion and balance control are based on physical models of vehicle dynamics. There are well developed vehicle dynamics models for four-wheel ground moving (e.g.,~\cite{Arab2021Acc,Bianchi20117174Active,Kabzan2019Racing,Seo2022Vehicle}). One challenge for autonomous ski-stunt maneuvers is the lack of accurate vehicle dynamics model. Although there are extensive research that study the rollover sequence and rollover detection~\cite{Rollover2013Phanomchoeng,Lu2007,Ralf2000Rollover}, the dynamics models in these work cannot be directly used for ski-stunt maneuver control.

\begin{figure*}[ht]
\hspace{-2mm}
\vspace{-0mm}
  \subfigure[]{
  \label{Fig_Truck}
  \includegraphics[width=2.46in]{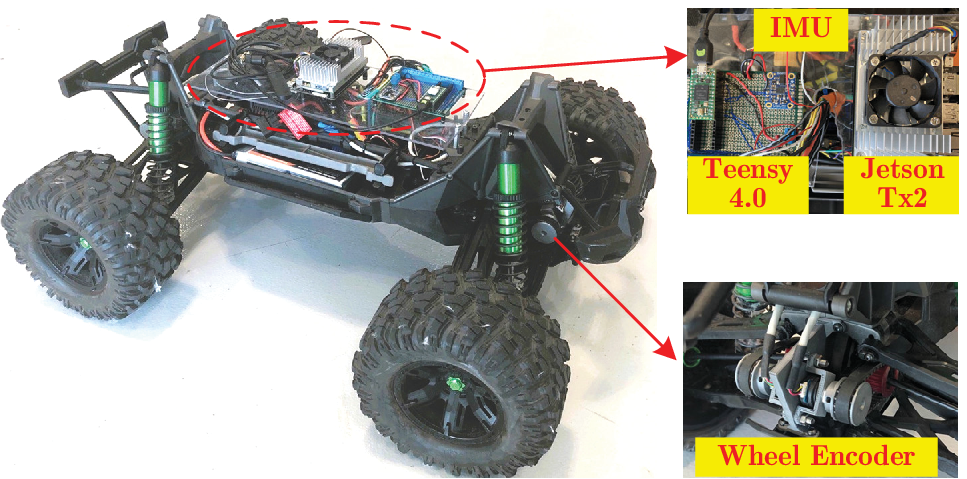}}
\hspace{-2mm}
  \subfigure[]{
	\label{Fig_Truck_Stunt}
	\includegraphics[width=1.93in]{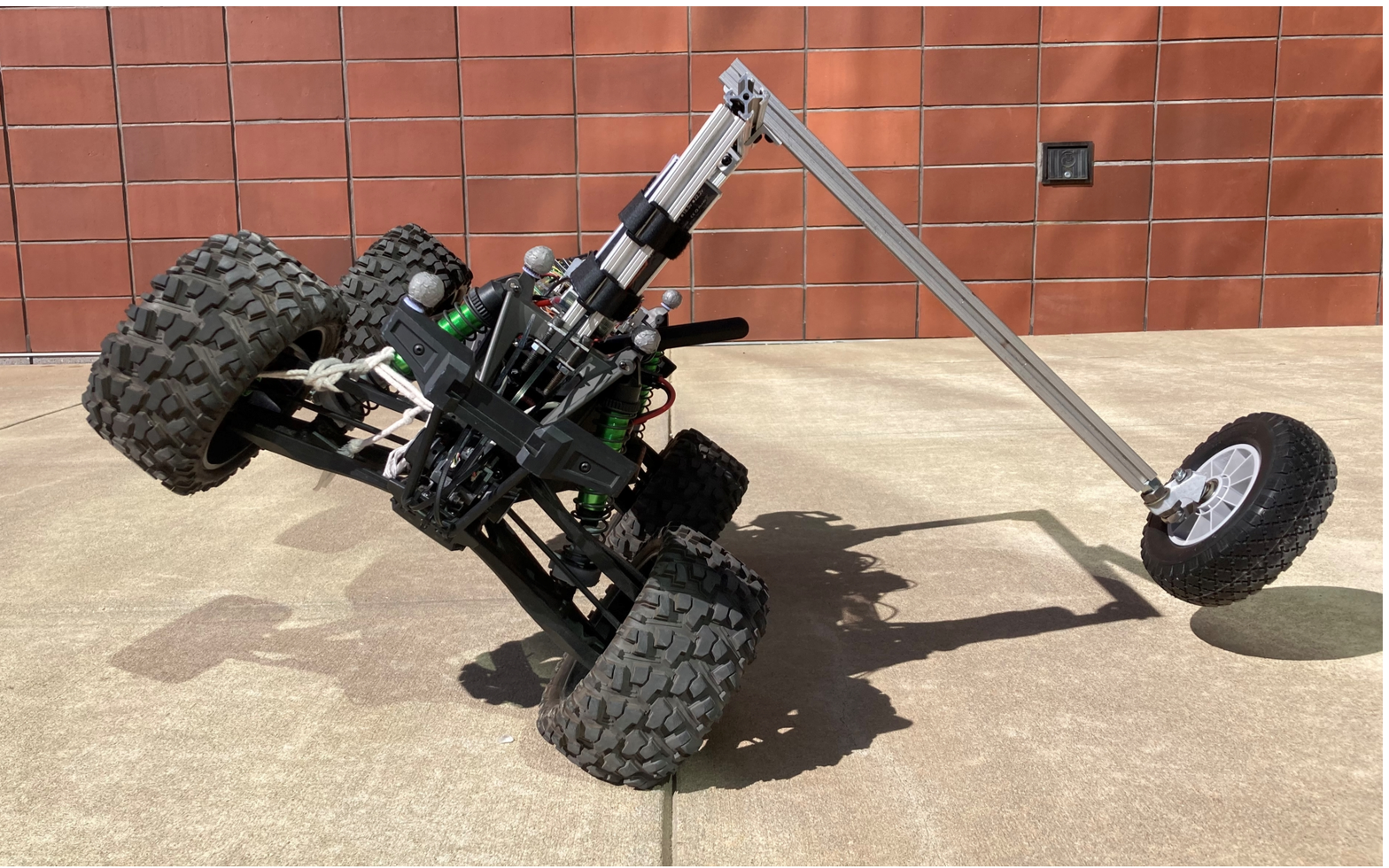}}
\hspace{-2mm}
   \subfigure[]{
   \label{Fig_Truck_Schematics_SideView}
  \includegraphics[width=1.45in]{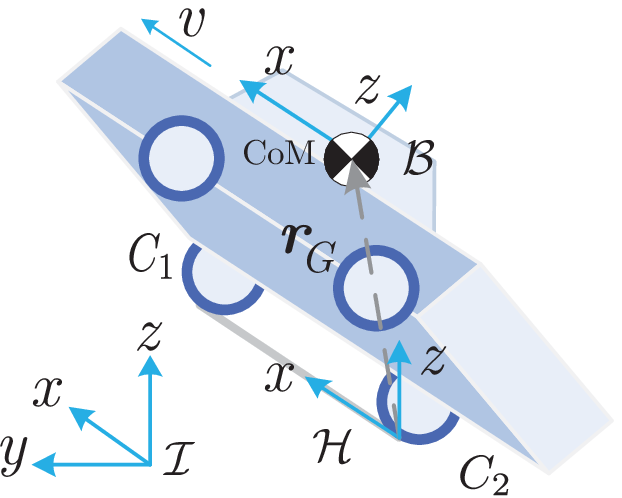}}
  \hspace{-5mm}
   \subfigure[]{
   \label{Fig_Truck_Schematics_BackView}
  \includegraphics[width=1.15in]{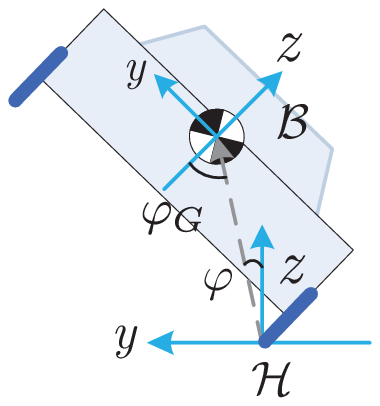}}
  \vspace{-1mm}
  \caption{(a) The scaled racing truck (modified from \href{https://traxxas.com/products/landing/maxx/} {Traxxas Maxx racing truck}) with Jetson TX2 computer and sensors. (b) The vehicle ski-stunt maneuver in experiment. (b) Side view and (d) back view of the schematics of the vehicle ski-stunt maneuver.}
  \label{Fig_Truck_Schematics}
  \vspace{-3mm}
\end{figure*}

In this paper, we design the safety-guaranteed motion control of autonomous ski-stunt maneuvers. Safety criteria for planar motion and roll motion are guaranteed for the collision avoidance and the roll balance when conducting ski-stunt maneuvers. To achieve superior performance, a machine learning-based data-driven model using Gaussian process (GP) regression is used to complement the nominal model and compensate for the modeling errors. A probabilistic exponential CBF is defined for planar motion safety by introducing the learning errors. The control design based on the nominal dynamics model is then extended through a model predictive control (MPC) by considering the safety requirement. To prevent any possible rollover, CBFs for roll motion are also added into the design. The BEM is introduced to estimate the desired instantaneous equilibrium~\cite{HanRAL2021, GetzPhD}. Finally, we update the control by regulating the roll motion to the BEM. Under the proposed control design, the closed-loop system is exponentially stable and the safety are guaranteed. We validate and demonstrate the design using a scaled trunk platform with simulation and experimental results. The main contribution of this work lies in the  proposed new learning-based safety and balance control for a ski-stunt maneuver. To our best knowledge, no autonomous ski-stunt maneuver has been reported in literature and this is the first study to demonstrate the safety-guaranteed ski-stunt maneuver. The proposed control method can also be applied to other underactuated robots, such as autonomous bicycle~\cite{WangTASE2019, ChenIJRR2015} and autonomous safely agile driving~\cite{Arab2021Acc,Arab2021Instructed} .

The remainder of the paper is outlined as follows. In Section~\ref{Sec_System Dynamics}, we build a machine learning-enhanced vehicle dynamics model using GP regression. The safe control design is presented in Section~\ref{Sec_Safe_Control}. We present the simulation and experimental results in Section~\ref{Sec_Result}. Finally, we summarize the concluding remarks in Section~\ref{Sec_Conclusion}.

\section{Learning-Enhanced Vehicle System Dynamics}
\label{Sec_System Dynamics}

In this section, we first present a nominal model of the vehicle dynamics. To improve the model accuracy, a GP-based enhancement is constructed for both the planar and roll vehicle motion.

\subsection{System Configuration and Problem Statement}

Fig.~\ref{Fig_Truck} shows the scaled truck platform that is modified for the ski-stunt maneuvers, while Fig.~\ref{Fig_Truck_Stunt} shows a snapshot of ski-stunt maneuver experiment. Figs.~\ref{Fig_Truck_Schematics_SideView} and~\ref{Fig_Truck_Schematics_BackView} show the side and back views of the vehicle under ski-stunt maneuver. The wheel contact line is denoted as $C_1C_2$ with front and rear wheel contact points as $C_1$ and $C_2$, respectively. Three coordinate frames are setup and used: The initial frame $\mathcal{I}$ is fixed on the ground with upward $z$-axis, the body frame $\mathcal{B}$ is fixed at the center of mass (CoM) of the truck and is used to describe the roll motion, and the local frame $\mathcal{H}$ is located at $C_2$ with the $x$-axis along $C_1C_2$ and obtained from $\mathcal{I}$ by rotating about the $z$-axis with vehicle yaw angle. We denote the position of $C_2$ as $\bm r_{C_2}= [x \, y]^T$ in $\mathcal{I}$ and $\bm r_{C_2}$ is considered the vehicle position. The steering, yaw and roll angles are denoted as $\phi$, $\psi$ and $\varphi$, respectively. The vehicle CoM position is denoted as $\bm r_G=[x_G \,y_G \,z_G]^T$ in $\mathcal{H}$ and the wheelbase is denoted as  $l_1=\overline{C_1C_2}$.

The vehicle's longitudinal velocity is denoted as $v$. The two inputs are $v$ and steering angle $\phi$, while the vehicle motion has three DOFs, i.e., planar motion $\bm r_{C_2}$ and roll motion $\varphi$. The zero roll angle $\varphi = 0$ is defined as the static equilibrium point, i.e., the CoM projection point on the ground located on $C_1C_2$ line. It is clear from Fig.~\ref{Fig_Truck_Schematics_BackView} that the position of $\varphi = 0$ corresponds to that rotating the vehicle $\varphi_G= \frac{\pi}{2}-\arctan|\frac{z_G}{y_G}|$ along the $x$-axis in $\mathcal{H}$ from regular four-wheel driving case. The actual rotation angle from the four-wheel driving position is then $\varphi_r=\varphi+\varphi_G$.

{\em Problem Statement}: The main task in this study is to design a vehicle controller to follow a desired trajectory and avoid the any obstacles while maintain the balance (preventing any possible rollovers) when running on two one-side wheels at $C_1$ and $C_2$ (i.e., ski-stunt maneuver).

\subsection{System Dynamics}

We first build a nominal dynamics model of the vehicle for planar and roll motion. Assuming no wheel-slippage conditions at $C_1$ and $C_2$ with the nonholonomic constraint at $C_2$, the vehicle velocity relationship is given as
\begin{equation}\label{Eq_Velo}
 \dot{x} = v \cos \psi=v \c_\psi,\; \dot{y} = v\sin \psi=v\s_\psi,
\end{equation}
where notations $\c_\psi=\cos \psi$ and $\s_\psi=\sin \psi$ are used for $\psi$ and other angles. The planar motion kinematic model is obtained by taking differentiation of~(\ref{Eq_Velo})
\begin{equation}\label{Eq_Planar_Kine}
\begin{bmatrix}
  \ddot x \\ \ddot y \end{bmatrix} = \underbrace{\begin{bmatrix}  \c_\psi & - v\s_\psi \\ \s_\psi & v\c_\psi \end{bmatrix}}_{\bm g_r(v,\psi)}
\begin{bmatrix} \dot v \\ \dot \psi \end{bmatrix} =\bm g_r(v,\psi)\bm u,
\end{equation}
where $\bm u =[u_v\; u_\psi]$, $u_v=\dot{v}$, and $u_\psi=\dot{\psi}$ is the yaw angle rate that is related to the steering control input as~\cite{WangTASE2019}
\begin{equation}\label{Eq_Yaw_Rate}
    \dot{\psi}=\frac{v}{l_1 \c_{\varphi_r}} \tan \phi.
\end{equation}

The roll motion of the vehicle is captured by an inverted pendulum model using the Lagrangian method. The position of CoM in $\mathcal{I}$ is
\begin{equation}\label{Eq_COM_Position}
  \bm r_G^{\mathcal I} = {\bm r}^\prime + \bm R_{\mathcal H}^{\mathcal I} \bm r_G^{\mathcal H}
  =\begin{bmatrix}  \bm r_{C_2} \\  0\end{bmatrix}+\begin{bmatrix}
x_{G} \c_\psi+l_G \s_\psi \s_\varphi \\
 x_{G} \s_\psi-l_G \c_\psi \s_\varphi
 \\ l_G \c_\varphi
\end{bmatrix},
\end{equation}
where  $\bm r^\prime=[\bm r_{C_2}^T\; 0]^T$, $\bm R_{\mathcal H}^{\mathcal I} \in SO(3)$ is the rotational matrix that transfers the vector in $\mathcal H$ to $\mathcal I$, $\bm r_G^\mathcal{H}=[x_G\; -l_G \s_\varphi\; l_G \c_\varphi]^T$, and $l_G =\sqrt{y_G^2+z_G^2}$. The velocity of the CoM in $\mathcal I$ is obtained by taking differentiation of~(\ref{Eq_COM_Position}), $\bm v_G^{\mathcal I} = \dot{\bm r}_G^{\mathcal I} $. The angular velocity of the vehicle in $\mathcal{B}$ is
\begin{equation*}
  \bm \omega_G^\mathcal{B}= R_\mathcal{H}^{\mathcal B}\bm \omega_G^\mathcal{H},\;\bm \omega_G^\mathcal{H}=[\dot\varphi\; 0\; \dot{\psi}]^T.
\end{equation*}

The kinetic energy of the vehicle is $T=\frac{1}{2}m({ \bm v_G^I})^2 + \frac{1}{2}(\bm \omega_G^\mathcal{B})^T\bm J_{G}\bm \omega_G^\mathcal{B}$, where $\bm J_{G} = \diag(J_x, J_y, J_z)$ is the moment of inertia about CoM. The potential energy is $V= mgl_G\c_\varphi$, where $m$ is the vehicle mass. With the Lagrangian method, we obtain the roll motion dynamics
\begin{equation}\label{Eq_Roll_Dyna}
  J_t\ddot \varphi  - mgl_G\s_\varphi = \tau,
\end{equation}
where $J_t=mr_G^2 + J_x$ and the steer-induced torque is $\tau=m l_{G} x_{G} \ddot{\psi} \c_\psi +m l_{G}^{2}  \dot{\psi}^{2} \s_\varphi\c_\varphi+\left(\c_\varphi^2 J_z+\s_\varphi^2 J_{y}\right) \dot{\psi}^{2} -\left(m l_{G} x_{G} \dot{\varphi} \s_\varphi-v l_{G} m \c_\varphi\right) \dot{\psi}$. The above steer-induced torque $\tau$ captures the centrifugal force, which influences the roll motion of the vehicle~\cite{Tanaka2009}. Noting that $v \gg |\dot{\psi}|$ and by neglecting the second order term and substituting~(\ref{Eq_Yaw_Rate}) into the steering torque, we obtain the simplified steer-induced torque as
\begin{equation}
\label{Eq_Steering_Troque}
\tau =mvl_G\c_\varphi u_\psi  =\frac{mv^2l_G\c_\varphi}{l_1\c_{\varphi_r}}\tan \phi.
\end{equation}

Letting $f_\varphi=\frac{1}{J_t}(mgl_G\s_\varphi)$ and $g_\varphi=\frac{1}{J_t}mvl_G\c_\varphi$, the roll dynamics in~(\ref{Eq_Roll_Dyna}) is re-written as
\begin{equation}\label{Eq_Roll_Dyna2}
  \ddot{\varphi}=f_\varphi+g_\varphi u_{\psi}.
\end{equation}
The nominal model of the entire system is written into the following compact form
\begin{equation}
	\label{Eq_Combined_Model}
\begin{bmatrix} \ddot{\bm r}_{C_2} \\ \ddot \varphi \end{bmatrix}= \underbrace{\begin{bmatrix}
  \bf 0 \\  f_\varphi \end{bmatrix}}_{\bm f} + \underbrace{\begin{bmatrix} \bm g_r(v,\psi) \\ \bm g_\varphi
\end{bmatrix}}_{\bm g }\begin{bmatrix}
   u_v\\
   u_\psi
\end{bmatrix} :=\bm f+\bm g \bm u,
\end{equation}
where $\bm g_\varphi=[0\; g_\varphi]$. It is clear that from~(\ref{Eq_Combined_Model}) the vehicle is underactuated and the steering input affects both the planar and roll motion.

\subsection{Learning-Enhanced Dynamics Model}

In~(\ref{Eq_Combined_Model}), the wheel slippery is neglected and the roll motion is captured by an approximated inverted pendulum model. Meanwhile, the vehicle platform is not ideally symmetric and rigid. To account for those unmodeled dynamics, we consider using the machine learning-based data-driven method to capture the actual system model from experimental data.

The actual model that captures the vehicle motion is modified and extended from~(\ref{Eq_Combined_Model}) as
\begin{equation}
	\label{Eq_Model_CoM}
  \ddot{\bm x} =\bm f+\bm g \bm u + \bm f_u,
\end{equation}
where $\bm x = [x\; y\; \varphi]^T=[{\bm r}^T_{C_2}\; \varphi]^T$, $\bm f_u =[f_{ux}\; f_{ux}\; f_{u\varphi}]^T \in \mathbb{R}^3$ denotes the unmodeled effects and system uncertainties. We assume that $\bm f_u$ is invariant of the vehicle's position ${\bm r}_{C_2}$. GP regression is used to capture the unmodeled dynamics. Assuming the dynamics $\bm f_u$ is related with the system state $\bm \xi = [\dot{x}\; \dot{y}\;  \ddot{x}\; \ddot{y}\; \varphi\; \dot{\varphi}\;  \ddot{\varphi}\; \phi\;  v\; ]^T \in \mathbb{R}^9$ as
\begin{equation}\label{Eq_Resudial}
  \bm f_w=\bm f_u(\bm \xi) + \bm w,
\end{equation}
where $\bm f_w$ denoted the noisy observation of $\bm f_u$ and $\bm w \sim \mathcal{N}(\bs 0,\bm \Sigma)$ is the zero-mean Gaussian noise. The training data set is $\bm D=\{\bm{X}, \bm{Y}\}=\left\{\{\bm \xi_i\}_{i=1}^N, \{\bm f_{wi}\}_{i=1}^N\right\}$, where $\bm f_{w}$ is obtained as the difference between actual measurement and nominal model calculation, and $\bm{X}$ and $\bm{Y}$ also denote the matrices composed by $\bm \xi_i$ and $\bm f_{wi}$ respectively.

For multi-dimension output, GP regression is constructed in each dimension. The GP regression for $f_{ux}$, for instance, is to maximize the likelihood function
\begin{equation}\label{Eq_GP_Kernel}
  \log(\bm Y_x;\bm X, \bm \theta) = -\frac{1}{2} \bm{Y}_x^{T}\bm{K}^{-1} \bm{Y}_x  -\frac{1}{2}\log\det(\bm K)
\end{equation}
where $K_{ij}=k(\bm\xi_i, \bm\xi_j)=\sigma_{f}^{2} \exp (-\frac{1}{2}(\bm\xi_i-\bm\xi_j)^{T} \bm{W}(\bm\xi_i-\bm\xi_j))+\sigma_0^2\delta_{ij}$, $\bm W = \diag\{W_1,\cdots, W_m\}$, $\delta_{ij}=1$ for $i=j$ only, $\bm \theta =\{\bm W, \sigma_f, \sigma_0\}$,  and $\bm{Y}_x$ is the vector composed by all $f_{wx}$ in $\bm D$. Given the new measurement data $\bm \xi^*$, GP model predicts the mean value and the standard deviation of the unmodeled dynamics as
\begin{align}
\mu(\bm \xi^*) = \bm{k}^{T}\bm{K}^{-1} \bm Y_x, \,\Sigma(\bm \xi^*)=k^{*}-\bm{k}^{T}\bm{K}^{-1} \bm{k},
\label{Eq_GP_Pred}
\end{align}
where $\bm{k}=\bm k(\bm \xi^*, \bm X)$ and $k^*= k(\bm \xi^*, \bm \xi^*)$. The predictions for $f_y$ and $f_\varphi$ are obtained in the same manner. We then use the prediction $\bm f_\mu(\bm \xi) = [f_{\mu x}(\bm \xi)\; f_{\mu y}(\bm \xi)\; f_{\mu \varphi}(\bm \xi) ]^T$ to approximate $\bm f_u$. Furthermore, the prediction error $\bm \delta_f =\bm f_u(\bm \xi) -\bm f_\mu(\bm \xi)$ is bounded in the sense of probability as shown in the following emma.
\begin{lem}[\hspace{-0.1mm}\cite{BECKERS2019390}]
 Given the training dataset $\bm D$, if the kernel function~(\ref{Eq_GP_Kernel}) is chosen such that $f_{ui}$ has a finite reproducing kernel Hilbert space norm $\left\|\bm{f}_u\right\|_{\bm{k}}<\infty$, for given $0<\eta<1$,
\begin{equation}\label{Eq_GP_Error}
  \mathrm{Pr} \left\{\|\bm \delta_f \| \leq\|\bm \Sigma ^{\frac{1}{2}}(\bm \xi) \bm{\kappa}\|\right\} \geq \eta,
\end{equation}
where $\mathrm {Pr}(\cdot)$ denotes the probability of an event, $\eta \in (0,1)$, $\bm \kappa, \bm \varsigma \in \mathbb{R}^{m}$, $\kappa_{i}=\sqrt{2| f_{ui}|_{\bm k}^{2}+300 \varsigma_{i} \ln ^{3} \frac{N}{1-\eta^{\frac{1}{m}}}}, \quad \varsigma_{i}=\max _{\bm \xi, \bm \xi^{\prime} \in \bm{X}} \frac{1}{2} \ln | 1 +\sigma_{i}^{-2} k_i\left(\bm \xi, \bm \xi^{\prime}\right) |$, and $i=1,2,3$ for the dimensional elements of $\bm f_u$.
\end{lem}

With the above discussion, the GP-based learning-enhanced vehicle dynamics model is obtained from~(\ref{Eq_Model_CoM}) as
\begin{equation}
\label{Eq_Hybrid_Model}
\dot {\bs \chi} = \underbrace{\begin{bmatrix} \dot{\bs{x}} \\ \bs{f} \end{bmatrix}}_{\bm F} +\underbrace{\begin{bmatrix} \bs{0} \\ \bs{f}_u \end{bmatrix}}_{\bm F_\mu}+\underbrace{\begin{bmatrix} \bs{0} \\ \bs{g} \end{bmatrix}}_{\bm G} \bm u +\underbrace{\begin{bmatrix} \bs{0} \\ \bs{\delta}_f \end{bmatrix}}_{\bm \delta_{F}}=\bm F+\bm F_\mu+\bm G \bs{u}+\bm \delta_{F},
\end{equation}
where ${\bs \chi} = [\bm x^T\; \dot{\bm x}^T]^T$, $\bm F_{\mu}$, $\bm F$, $\bm G$ and $\bm \delta_{F}$ are in proper dimensions.

\section{Safe Ski-Stunt Maneuver Control}
\label{Sec_Safe_Control}

In this section, we design the safe vehicle control strategy. Both the planar and roll motion safety (rollover prevention and balance maintaining) are considered. The probabilistic exponential CBF is introduced to deal with the maneuver safety, while the roll motion is stabilized on the BEM to guarantee the balance.

\subsection{Control Barrier Function with Learning Model}

For ski-stunt motion, we design learning-based CBF control strategy to prevent possible collisions. The interpretation of the safety is that the state of the vehicle dynamics remains within a safety set defined by
\begin{equation}
\label{Eq_Safe_Set}
    \mathcal S  =\{\bm \chi: h(\bm \chi) \geq 0\},
\end{equation}
where $h(\cdot)$ is the continuous differential function. Set $\mathcal{S}$ is referred as a safety set of the vehicle motion. We assume that $h(\bm \chi)$ has the relative degree $p$, that is, the control input $\bm u$ appears in the $p^\mathrm{th}$ derivative of $h(\bm \chi)$. To consider the general case of the safety requirement, the explicit form of $h(\bm \chi)$ is not specified here.

To define the CBF for the nonparametric model~(\ref{Eq_Hybrid_Model}), we introduce variable $\bs{q} \in \mathbb{R}^p$ in terms of the function $h(\bm \chi)$ as
\begin{equation}
  \bm q(\bm \chi)=
  \begin{bmatrix}
    h(\bm \chi)\\
    \vdots\\
    h^{(p-1)}(\bm \chi)\\
  \end{bmatrix}
  =
  \begin{bmatrix}
  h(\bm \chi)\\
  \vdots\\
  L_F^{p-1} h(\bm \chi)\\
  \end{bmatrix}
  \end{equation}
and Lie derivative $L_F h(\bm \chi) = L_{F}h(\bm \chi) +L_{F\mu}h(\bm \chi)$. The dynamics of $\bm q$ is
\begin{equation}\label{Eq_sys1}
  \dot{\bm q} = \bm A \bm q+ \bm B \bm u_h + \bm u_\delta, h = \bm C \bm q
\end{equation}
with
\begin{equation*}
  \bm A = \begin{bmatrix}
  0&\bm I_{p-1} \\
  \bs 0&\bs 0
\end{bmatrix},
  \bm B=\begin{bmatrix}
  \bs 0 \\
  1
\end{bmatrix},
 \bm C= \begin{bmatrix}
  1&\bs 0
\end{bmatrix},
\bm u_\delta = \begin{bmatrix}
\bs 0\\
\frac{\partial h}{\partial \bm \chi} \bm \delta_{F}
\end{bmatrix}
\end{equation*}
where $\bm I_{p-1}$ represents the identity matrix of dimention $p-1$, $u_h=L_F^ph(\bm \chi)+L_GL_F^{p-1}h(\bm \chi) \bm u$, and $L_GL_F^{p-1}h(\bm \chi)$ is assumed to be invertible.

The feedback gain $\bm \gamma$ is selected properly such that the unperturbed system $(\bm A, \bm B, \bm C)$ ($\bs{u}_\delta=\bs{0}$) with control input $\bs{u}_h=-\bm \gamma \bm q$ is exponentially stable. The solution of $h(\bm \chi )$ then becomes
\begin{align}
  h(\bm \chi)&= \underbrace{\bm{C}e^{-(\bm{A} - \bm{B\gamma})t}\bm{q}(0)}_{h_u(\bm \chi)}+ \underbrace{\int_0^t \bm{C}e^{(\bm{A} - \bm{B\gamma})\nu} \bm u_\delta(t - \nu)d\nu}_{h_\delta(\bm \chi)},\nonumber\\
  &=h_u(\bm \chi)+h_\delta(\bm \chi),
\end{align}
If the model is exactly accurate, that is, $h_\delta(\bm \chi)=0$, $h(\bm \chi)$ is referred as the exponential CBF, when $\bm u_h > -\bm \gamma \bm q$ and then $h_u(\bm \chi) \ge \bm{C}e^{(\bm{A} - \bm{B\gamma})t}\bm{q}(0)>0$, for $t>0$ and $h_u(0)>0$~\cite{Ames2019Barrier, Nguyen2016Exponential}. Assuming that $h(\bm \chi)$ is locally Lipschitz in $\bm \chi \in \mathcal{S}$, namely, $\|\tfrac{\partial h}{\partial \bm \chi}\|\le M_h$ with finite number $M_h>0$, the term $\bm u_\delta$ can be shown as probabilistically bounded  $\Pr \left\{\| \bm u_\delta\| \leq {M_h}\|\bm \Sigma ^\frac{1}{2}(\bm \xi )\bm \kappa\|  \right\} \geq \eta$. Then $h_\delta (\bm \chi)$ is bounded with probability
\begin{equation*}
  \Pr \left\{ | h_\delta  | \leq h_\delta^{\max } \right\} \geq \eta,
\end{equation*}
where $h_\delta^{\max} =\sup_{\{\bm{\delta}_{F}, t\}}\int_0^t \bm C e^{(\bm A -\bm B \bm \gamma ) \nu}\bm u_\delta(t-\nu) d \nu $ for all possible GP prediction errors.

\begin{defn}
Probabilistic exponential CBF: Given the nonparametric dynamics~(\ref{Eq_Hybrid_Model}), function $h(\bm \chi)$ is a probabilistic exponential CBF if there exists $\bm \gamma$ such that
  \begin{equation}\label{Eq_CBF1}
   \sup_{\bs{u} \in \mathcal{U}}\left[L_F^ph(\bm \chi)+L_GL_F^{p-1}h(\bm \chi) \bm u \right] \ge -\bm \gamma\bm q(\bm \chi),
  \end{equation}
where $\mathcal{U}$ is the set that contains all feasible control $\bs{u}$ and
  \begin{equation}\label{Eq_CBF2}
    h(\bm \chi)  =h_1(\bm \chi)-h_2(\bm \Sigma(\bm \xi))
  \end{equation}
with $h_1(\bm \chi)$ denoting the nominal function and $h_2(\bm \Sigma(\bm \xi))$ being introduced to account for the GP prediction uncertainties. $h_2(\bm \Sigma(\bm \xi))$ is chosen as $h_2(\bm \Sigma(\bm \xi)) = \bm \Sigma(\bm \xi)$~\cite{Khan2021Safety}. Meanwhile, if control $\bm u$ satisfies~(\ref{Eq_CBF1}), function $h(\bm \chi)$ has
  \begin{equation}\label{Eq_CBF3}
    \Pr \left\{h(\bm \chi) \ge -h_\delta^{\max}\right\} \ge \eta.
  \end{equation}
\end{defn}

Comparing with the conventional CBF, $h(\bm \chi)$ might reach to a negative value. However, with sufficient training data, GP prediction error is small~\cite{BECKERS2019390} and therefore, $h_\delta^{\max} \ll 1$. In practice, a safety buffer zone can be added when designing the nominal CBF by considering the vehicle size, which is interpreted as to define the safety criterion from a conservative perspective. Furthermore, the CBF in~(\ref{Eq_CBF3}) incorporates the probability property of GP regression, which is not considered in other CBF control works~\cite{Khan2021Safety,Seo2022Vehicle}.

\subsection{Ski-Stunt Maneuver Control}

With the CBF designed above, the ski-stunt maneuver control can be formulated in a safety critical control form~\cite{Ames2019Barrier}. The safety critical control does not explicitly design the control input and instead, the CBF is employed as a constraint to modify the nominal control. The set of safety guaranteed control is defined as
\begin{equation}
\label{Eq_Safe_Control_Set}
   \mathcal {U}_s=\{\bm u\in\mathcal{U}: L_F^ph(\bm \chi)+L_GL_F^{p-1}h(\bm \chi) \bm u \ge  -\bm \gamma\bm q(\bm \chi)\}.
\end{equation}
The control $\bm u$ is further modified to guarantee safety by solving the following programming problem~\cite{pmlr-v120-taylor20a}
\begin{align}
\mathop {\min }\limits_{\bm u_s \in \mathcal{U}} &  \bm e_u ^T \bm e_u, \; \text{subject to:} \;\; u_s\in \mathcal{U}_s,
\label{Eq_CBF_Control}
\end{align}
where $\bm e_u =\bm u_s-\bm u$, $\bm u$ is the nominal control and $\bm u_s =[u_{sv}\; u_{s\psi}]^T$ is the target safe control.

The safety criteria is the collision avoidance with multiple obstacles for the planar motion. For roll motion, to initialize the ski-stunt maneuvering from four-wheel driving mode, a large torque is needed to counter the gravity effect. We design a second CBF $h(\bm \chi)=h(\varphi)$ in terms of the roll motion to safely drive roll angle to desired profile and prevent a complete rollover. We take above planar motion safety and rollover prevention into the control design. To further strengthen the safety, given the GP prediction errors and the barrier function~(\ref{Eq_CBF2}), we extend~(\ref{Eq_CBF_Control}) and formulate the planar motion safe control design as a MPC problem as
\begin{subequations}
\label{Eq_MPC_BEM_CBF}
\begin{align}
    &\mathop{\min }\limits_{\bs{u}_H}   \int_t^{t + t_H} {\bm e^T}{\bm W_1}\bm e + \bm e_u ^T \bm W_2 \bm e_udt  \label{Eq_MPC_obj},\\
    \text{\hspace{-4mm} subj. to}:\,&\ddot {\bm x} =\bm f_{\mu} + \bm f + \bm g \bm u_s, \\
    &L_F^ph_i(\bm r)+L_GL_F^{p-1}h_i(\bm r) \bm u_s \ge -\bm \gamma_i\bm q_i(\bm r),\\
    &L_F^ph_j(\varphi)+L_GL_F^{p-1}h_j(\varphi) \bm u_s \ge -\bm \gamma_j\bm q_j(\varphi),
\end{align}
\end{subequations}
where $\bs{u}_H=\{\bm u_{s1},\cdots,\bm u_{sH}\}$ is the $H$-step predictive control input set, $t_H=H \Delta t$ is the prediction horizon, $\Delta t$ is the step length, and $H \in \mathbb{N}$. $\bm e=\bm \chi_d-\bm \chi$, $\bm \chi_d$ is the desired state, $\bm e_u=\bm u_s-\bm u$, and $\bm W_1\in\mathbb{R}^6, \bm W_2\in\mathbb{R}^2$ are positive definite diagonal matrices, $h_i(\bm r)$ is the $i$th CBF for planar motion and $h_j(\varphi)$ is the $j$th CBF for rollover prevention. $h_i(\bm r)$ and $h_j(\varphi)$ are defined in the same way as $h(\bm \chi)$ with other elements being zeros. The optimization problem is solved online in real time via gradient descending algorithm, such as sequential quadratic programming~\cite{nocedal2006sequential}.

The safe planar control strategy $\bm u_s$ in~(\ref{Eq_MPC_BEM_CBF}) does not necessarily guarantee the balance of the roll motion. To guarantee the stability and balance of roll motion, we first compute the BEM and the safe planar motion control are updated by embedding the roll motion around the BEM. Given the safe control input $\bm u_s=[u_{sv}\; u_{s\psi}]^T$, the BEM is defined as set of all instantaneous equilibrium roll angles, namely,
\begin{equation}
\label{Eq_BEM}
  \mathcal{E}=\{\varphi^e: f_\varphi(\varphi^e)+f_{\mu \varphi}(\varphi^e)+g_\varphi(\varphi^e)u_{s\psi}=0\}.
\end{equation}
We update the control input $\bm u_s$ by enforcing the roll motion moving around the BEM. Solving BEM requires to invert functions $f_\varphi$, $f_{\mu \varphi}$ and $g_\varphi$, which is time consuming for the learning model. Instead, we estimate the BEM by minimizing the following function
\begin{equation*}
 \min_{\varphi}   \{\Gamma(\varphi) = (f_{\varphi}(\varphi)+f_{\mu \varphi}(\varphi)+g_{\varphi}(\varphi) u_{s \psi})^2\}.
\end{equation*}
Furthermore, we solve the above optimization problem numerically by gradient descending procedures
\begin{subequations}
\label{Eq_BEM_Calculation}
  \begin{align}
   & \varphi_{i+1}^{e}=\varphi_{i}^{e}- \left.\alpha \frac{\partial \Gamma}{\partial \varphi}\right|_{\varphi_{i}^{e}},\\
   & \frac{\partial \Gamma}{\partial \varphi} =2 \Gamma\left(\frac{\partial f_{\varphi}}{\partial \varphi}+\frac{\partial g_{\varphi}}{\partial \varphi} u_{s \psi}+\frac{\partial f_{\mu \varphi}}{\partial \varphi}\right),\\
   & \frac{\partial f_{\mu \varphi}}{\partial \varphi} =\bm Y_{\varphi}^{T} \bm{K}^{-1}  \frac{\partial \bm{k}}{\partial \bm \xi} \frac{\partial \bm \xi}{\partial \varphi},
\end{align}
\end{subequations}
with $\bs{k}$ is given by the GP model estimate~(\ref{Eq_GP_Pred}), $\alpha>0$ and the iteration is terminated when $\Gamma (\varphi^e_i)\le \epsilon$ for $\epsilon>0$. The control input is finally updated by incorporating the BEM as
\begin{equation}
\label{Eq_Roll_Control}
  u_{\varphi\psi}=g_\varphi^{-1}(-f_\varphi-f_{\mu \varphi}+\ddot{\varphi}^e-k_{p}e_\varphi-k_{d}\dot{e}_\varphi)
\end{equation}
to enforce the roll motion moving around $\mathcal{E}$, where $e_\varphi=\varphi-\varphi^e$ and $k_{p},k_{d}>0$ are feedback gains. The final control then is $\bs{u}_s^f=[u_{sv}\; u_{\varphi\psi}]^T$.

\begin{algorithm}[h!]
	\SetKwData{Left}{left}
	\SetKwData{This}{This}
	\SetKwData{Up}{up}
	\SetKwFunction{Union}{Union}
	\SetKwFunction{FindCompress}{FindCompress}
	\SetKwInOut{Input}{Input}
	\SetKwInOut{Output}{Output}
	Specify $\bm W_1, \bm W_2, \bm \gamma_i,  \alpha, \epsilon, k_p, k_d, H$, $t=0$ and $t_H$\;
    Design the CBF function $h_i(\bm r)$ and $h_j(\varphi)$\;
    \While{$t<t_H$}
    {Design the nominal control $\bm u=[u_v\; u_\psi]^T$ by only considering the planar motion\;
    Solve~(\ref{Eq_MPC_BEM_CBF}) to obtain $\bm u_s=[u_{sv}\; u_{s\psi}]^T$\;
    Solve $\varphi^e$ by~(\ref{Eq_BEM_Calculation}) until $\Gamma (\varphi^e_i)\le \epsilon$\;
    Update the steering control $u_{\varphi\psi}$ by~(\ref{Eq_Roll_Control})\;
    Apply the control $\bs{u}_s^f=[u_{sv}\; u_{\varphi\psi}]^T$\;
    $t=t+\Delta t$\;}
	\caption{Safe control design for ski-stunt maneuver}
	\label{Algorithm_Safe_Control}
\end{algorithm}

Algorithm~\ref{Algorithm_Safe_Control} illustrates the overall safe control design for ski-stunt maneuver. In the algorithm, lines 4 to 7 can be interpreted as a online CBF-based trajectory planning, where a family of safety involved CBFs are considered as dynamic constraints for both the planar and roll motion. Line 8 represents the final control to the vehicle. The stability of the system with the learning model and the above control design is summarized in the following lemma, which follows directly from Theorem 1 in~\cite{HanRAL2021}.
\begin{lem}
\label{Lem_Truck_Stability}
Assuming that BEM estimation error (i.e., the GP regression error $\bm \delta_f$ and BEM estimation error in~(\ref{Eq_BEM_Calculation})) is locally Lipschitz and affine with the planar and roll motion errors $\bm e$. The system under control $\bs{u}_s^f$ is stable with high probability and the error of the closed system~(\ref{Eq_Hybrid_Model}) converges into a small ball around zero exponentially, that is, $\bm \chi \in \mathcal{S}$ and $\varphi \in \mathcal{E}$.
\end{lem}

\begin{figure*}[t!]
	\hspace{-5mm}
	\subfigure[]{
		\label{Fig_Diff_H_Traj}
		\includegraphics[width=4.8cm]{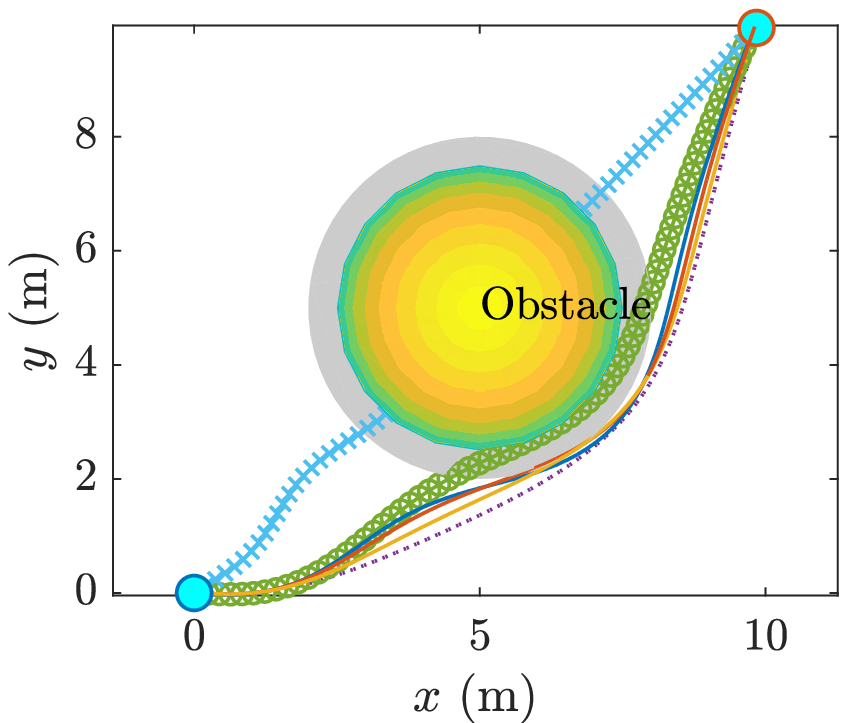}}
	\hspace{-6mm}
	\subfigure[]{
		\label{Fig_Diff_H_Roll}
		\includegraphics[width=4.6cm]{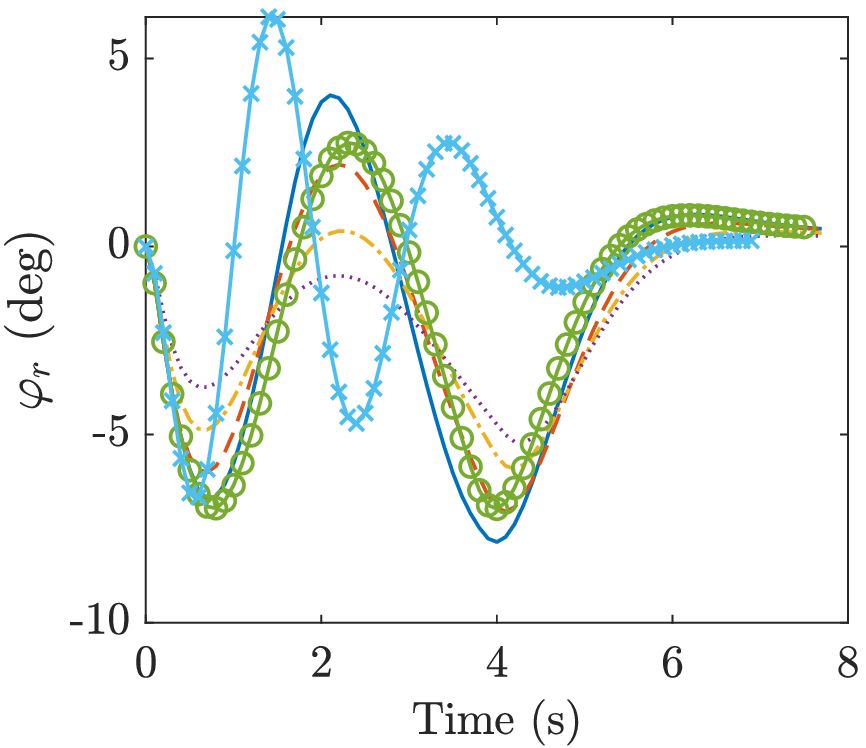}}
	\hspace{-6mm}
	\subfigure[]{
		\label{Fig_Diff_H_Steer}
		\includegraphics[width=4.6cm]{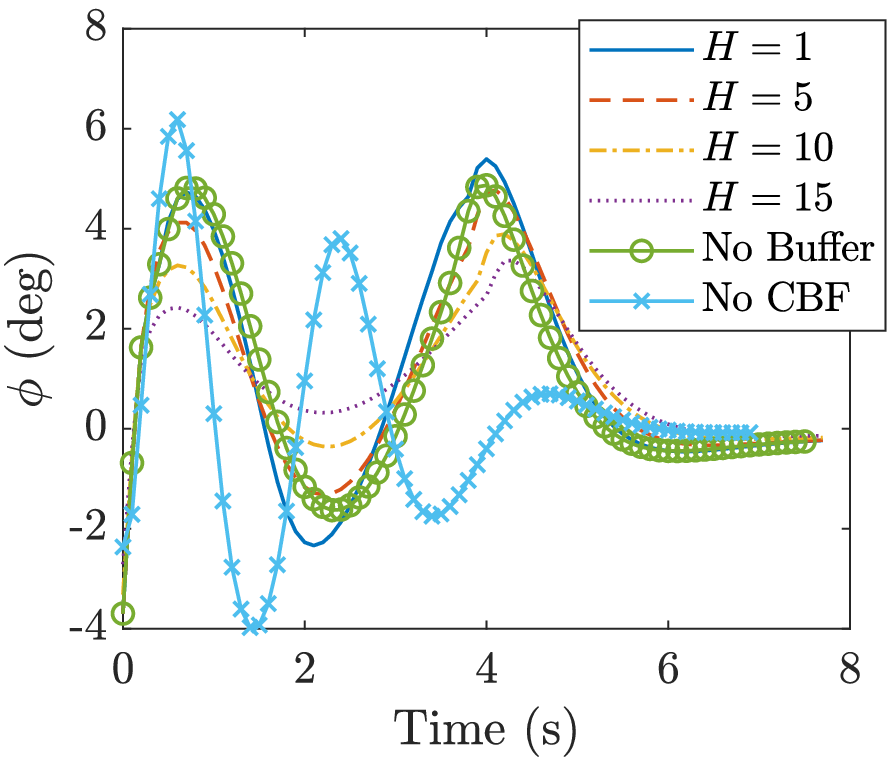}}
	\hspace{-4mm}
	\subfigure[]{
		\label{Fig_Diff_H_h}
		\includegraphics[width=4.7cm]{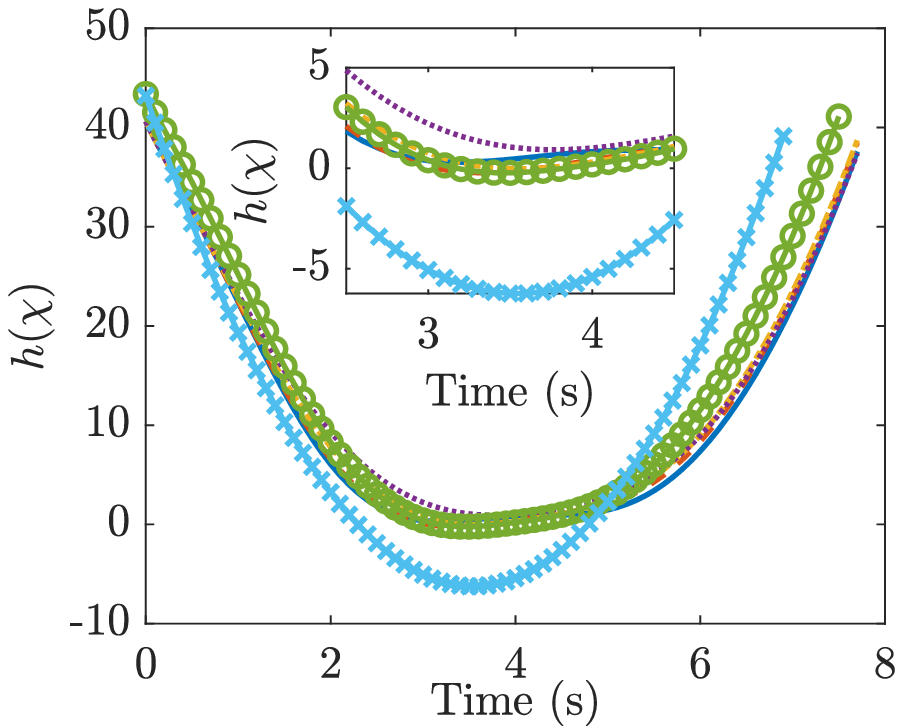}}
	\vspace{-4mm}
	\caption{Simulation results of truck ski-stunt maneuver with different prediction horizons. (a) Truck trajectory, (b) Roll angle, (c) Steering angle, (d) CBF value. The gray area in (a) denotes the buffer zone. The two circle markers shows the start and target locations.}
	\label{Fig_Diff_H}
	\vspace{-2mm}
\end{figure*}

\begin{figure*}[ht!]
\hspace{-4mm}
	\subfigure[]{
		\label{Fig_Diff_V_Traj}
		\includegraphics[width=4.7cm]{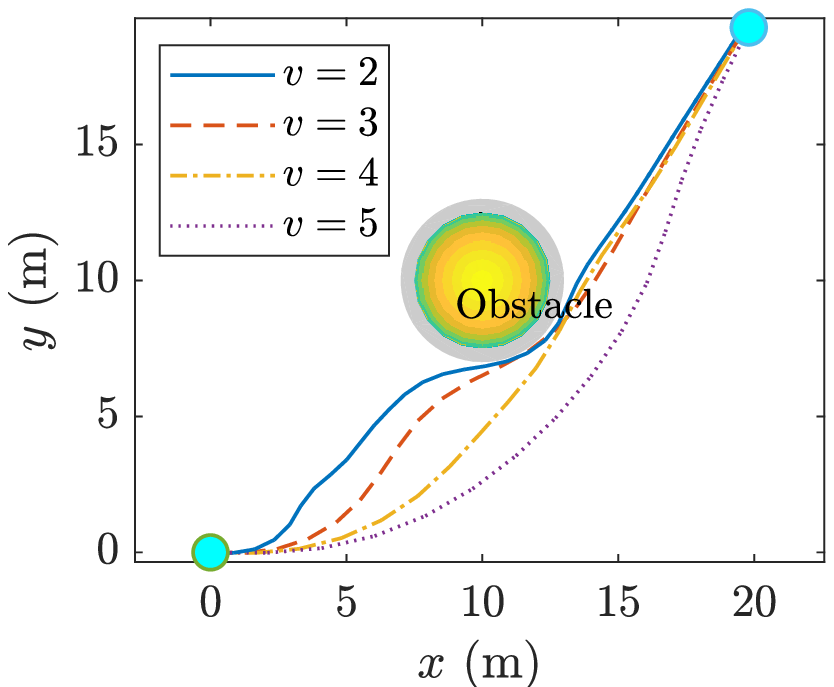}}
	\hspace{-6.5mm}
	\subfigure[]{
		\label{Fig_Diff_V_Roll}
		\includegraphics[width=4.7cm]{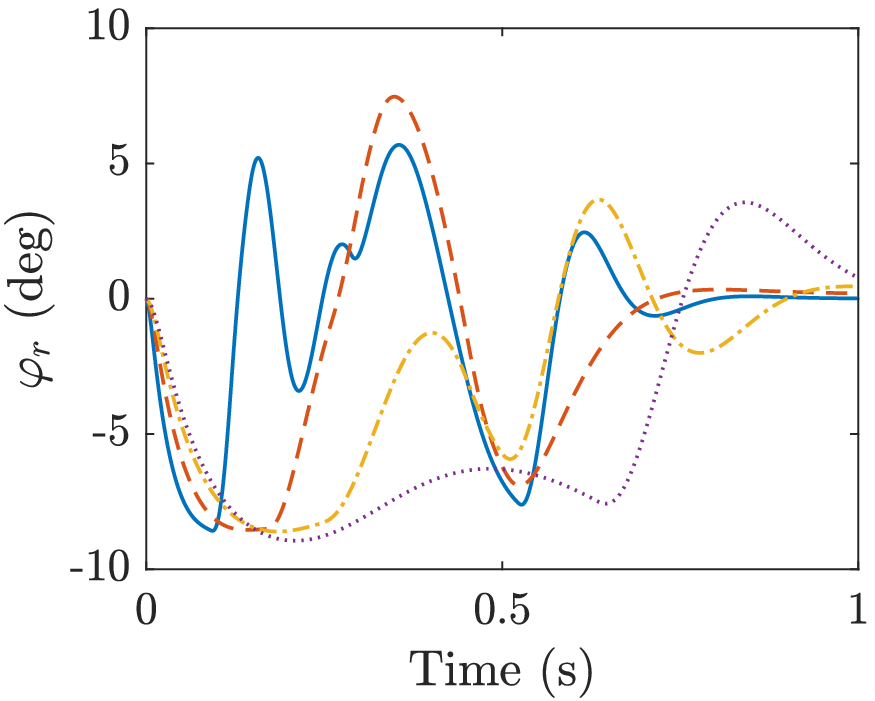}}
	\hspace{-6.5mm}
	\subfigure[]{
		\label{Fig_Diff_V_Steer}
		\includegraphics[width=4.7cm]{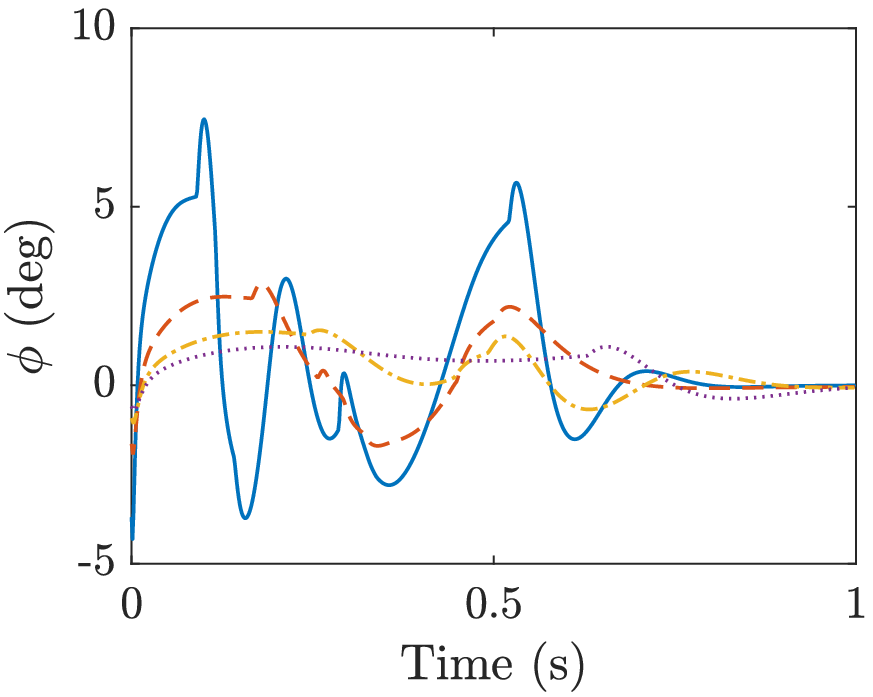}}
	\hspace{-6.5mm}
	\subfigure[]{
		\label{Fig_Diff_V_h}
		\includegraphics[width=4.8cm]{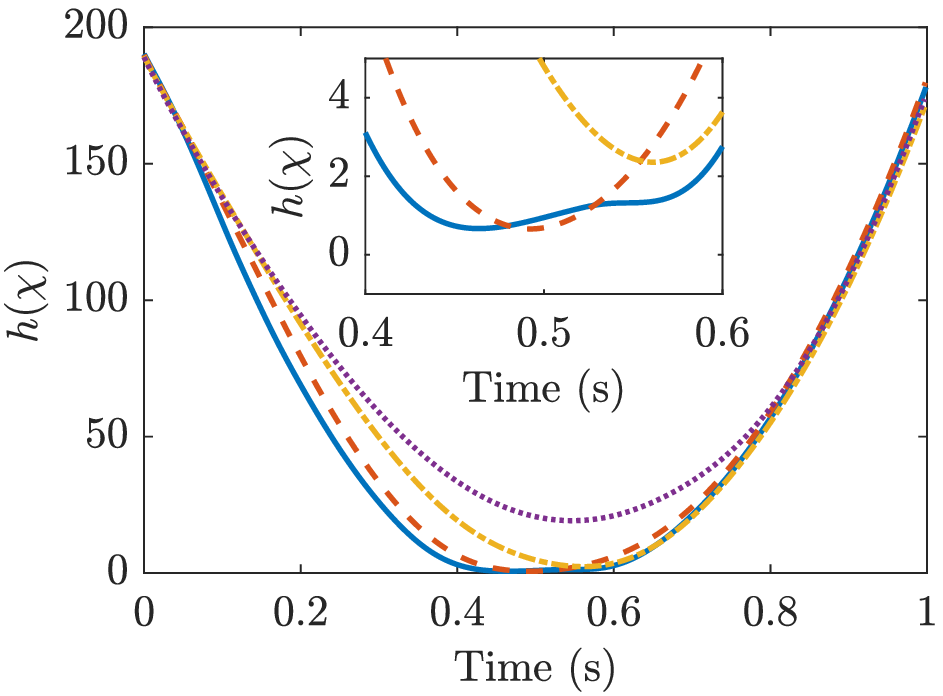}}
	\vspace{-3mm}
	\caption{Simulation results of truck ski-stunt maneuver with velocities. (a) Truck trajectory, (b) Roll angle, (c) Steering angle, (d) CBF value. The gray area in (a) denotes the buffer zone. The two circle markers shows the start and target locations. The time durations are normalized into $1$ s for all cases.}
	\label{Fig_Diff_V}
	\vspace{-2mm}
\end{figure*}

\section{Simulation and Experimental Results}
\label{Sec_Result}

We conduct simulation results and the vehicle model is based on the physical prototype as shown in Fig.~\ref{Fig_Truck}. Preliminary experiments are also included to demonstrate the feasibility of the control design.

\subsection{Experimental Setup}

The scaled racing truck was built and modified from an RC platform (model Maxx) from Traxxas. Encoders and inertia measurement unit (IMU) were installed to measure the front- and rear-wheel velocities and roll/yaw angles. A Jetson TX2 computer and Teensy 4.0 microcontroller were used for onboard computation purpose. Table~\ref{Table_Truck_Parameter} lists the values of the physical model parameters.

To conduct the simulation study, the unmodeled dynamics are considered as $f_{ux}=0.5v\c^2_\psi \s_\psi$, $f_{uy}=0.5v\c_\psi \s_\psi$ and $f_{u\varphi}=0.25v^2\s_\varphi - 0.25\dot \varphi$. In the nominal model, the moment of inertia of the vehicle was set as $J_t=1$ kg$\cdot$m$^2$, which is not accurate as the true value shown in Table~\ref{Table_Truck_Parameter}. The GP regression data were generated using the nominal model with arbitrarily designed input to excite the system. A total of 1000 data points were used for the GP model training. The obstacle in simulation had a circular shape with radius $R$. The vehicle in ski-stunt maneuver should avoid the obstacles while keeping balance. Starting from autonomous four-wheel driving to ski-stunt maneuvers, any possible rollovers should also be avoided. The CBFs were designed as
\begin{align*}
  h_i(\bm r)&=(R+R_\epsilon)^2-(x-x_{ci})^2-(y-y_{ci})^2-\Sigma_x -\Sigma_y,\\
  h(\varphi)&=(\varphi_{\max}+\varphi_G)^2-(\varphi+\varphi_G)^2-\Sigma_\varphi,\\
  h(\dot \varphi)&=\dot\varphi_{\max}^2-\dot \varphi^2,
\end{align*}
where $(x_{ci}, y_{ci})$ was the center position of the $i$th obstacle, $R_\epsilon$ was used to account for the GP regression error ($h_\delta$) and served as a buffer zone. $\varphi_{\max}$ was allowed maximum roll angle to prevent the rollover and $\dot\varphi_{\max}$ denoted the maximum roll angular velocity.

\renewcommand{\arraystretch}{1.3}
\setlength{\tabcolsep}{0.1in}
\begin{table}[h!]
	\centering
	\caption{Values for the model parameters of the scaled truck}
	\label{Table_Truck_Parameter}
	\vspace{-1mm}
	\begin{tabular}{|c|c|c|c|c|c|}
		\hline\hline
		$m$ (kg)& $J_t$ (kgm$^2$) & $l_1$ (m) & $y_G$ (m) & $z_G$ (m)& $\varphi_G$ (deg)  \\ \hline
		$11.4$ & $1.35$ & $0.48$ & $0.25$ & $0.29$ & 40\\	\hline\hline
	\end{tabular}
	\vspace{-1mm}
\end{table}

In simulation, the vehicle velocity was set as constant during the ski-stunt maneuver, which left the steering as the only control actuation for planar motion and roll motion. This limited actuation increases the control challenge. We compare the control performance at different velocities. The control parameters used in design were $\bm \gamma_i = [1\; 1.5]^T$, $k_p=35$, $k_d=20$,  $\bs{W}_1=\diag\{ 20, 20, 20, 10, 10, 10\}$, $\bs{W}_2=\diag\{5, 5\}$, $\alpha=0.05$, $\epsilon=0.005$, and $\Delta t=20$~ms. The computer platform used for simulation has a Core i7-9700 @ 3.0G Hz $\times$ 8 CPU.

\subsection{Simulation Result}

\setcounter{figure}{4}
\begin{figure*}[ht!]
\hspace{-4mm}
	\subfigure[]{
		\label{Fig_Track_Traj}
		\includegraphics[width=6.2cm]{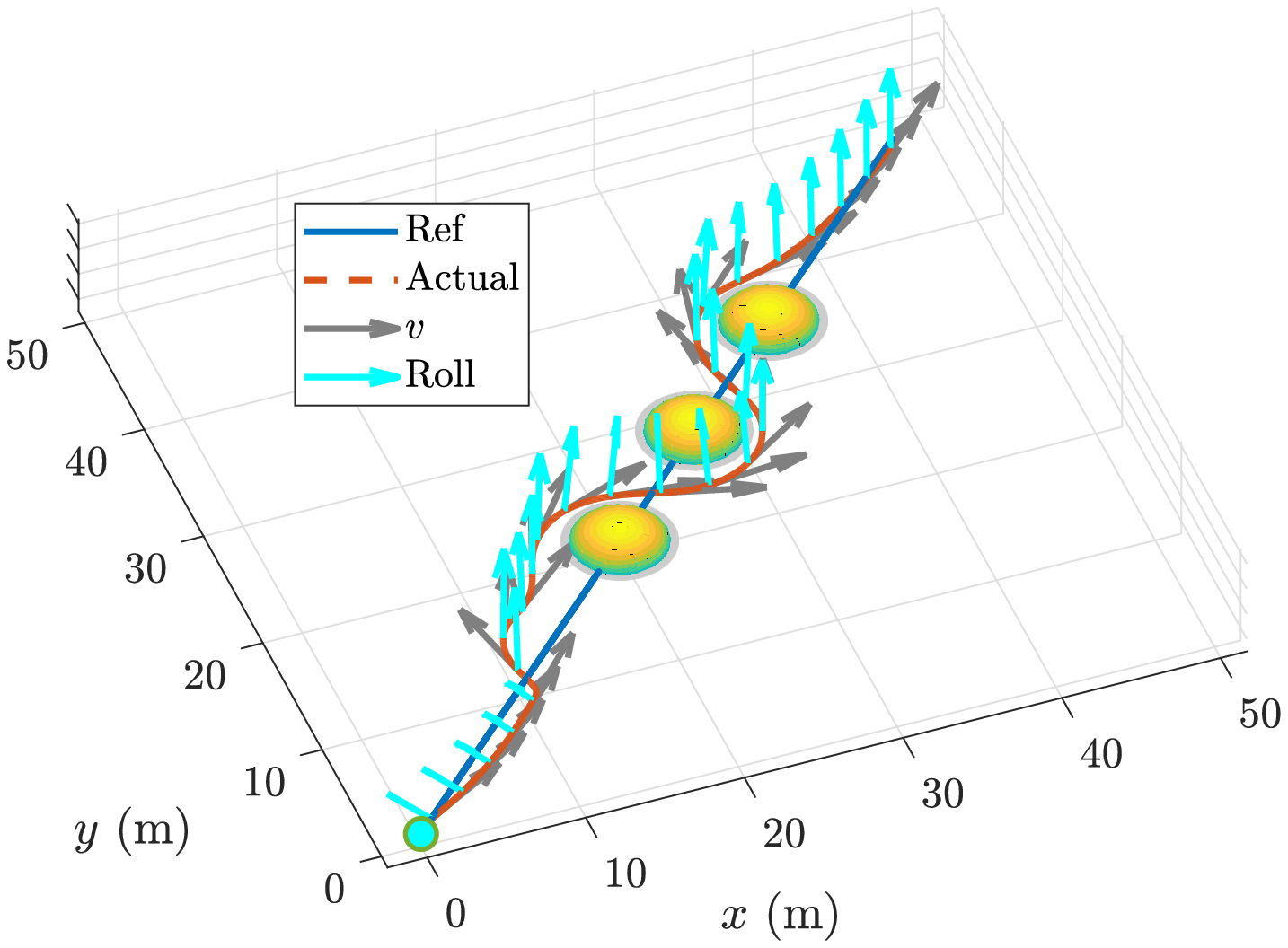}}
	\hspace{-4mm}
	\subfigure[]{
		\label{Fig_Track_Roll}
		\includegraphics[width=5.8cm]{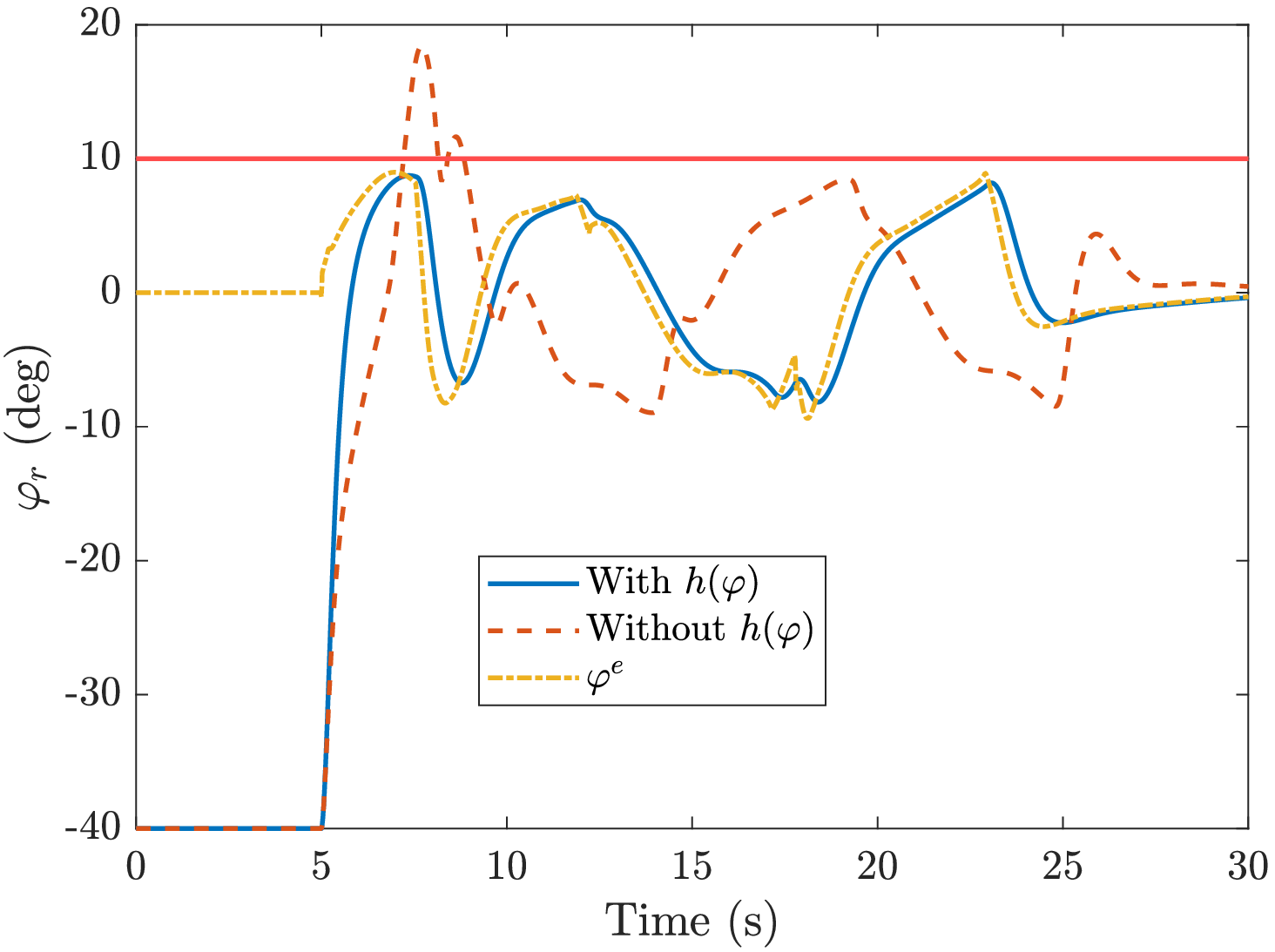}}
	\hspace{-4mm}
	\subfigure[]{
		\label{Fig_Track_h}
		\includegraphics[width=5.8cm]{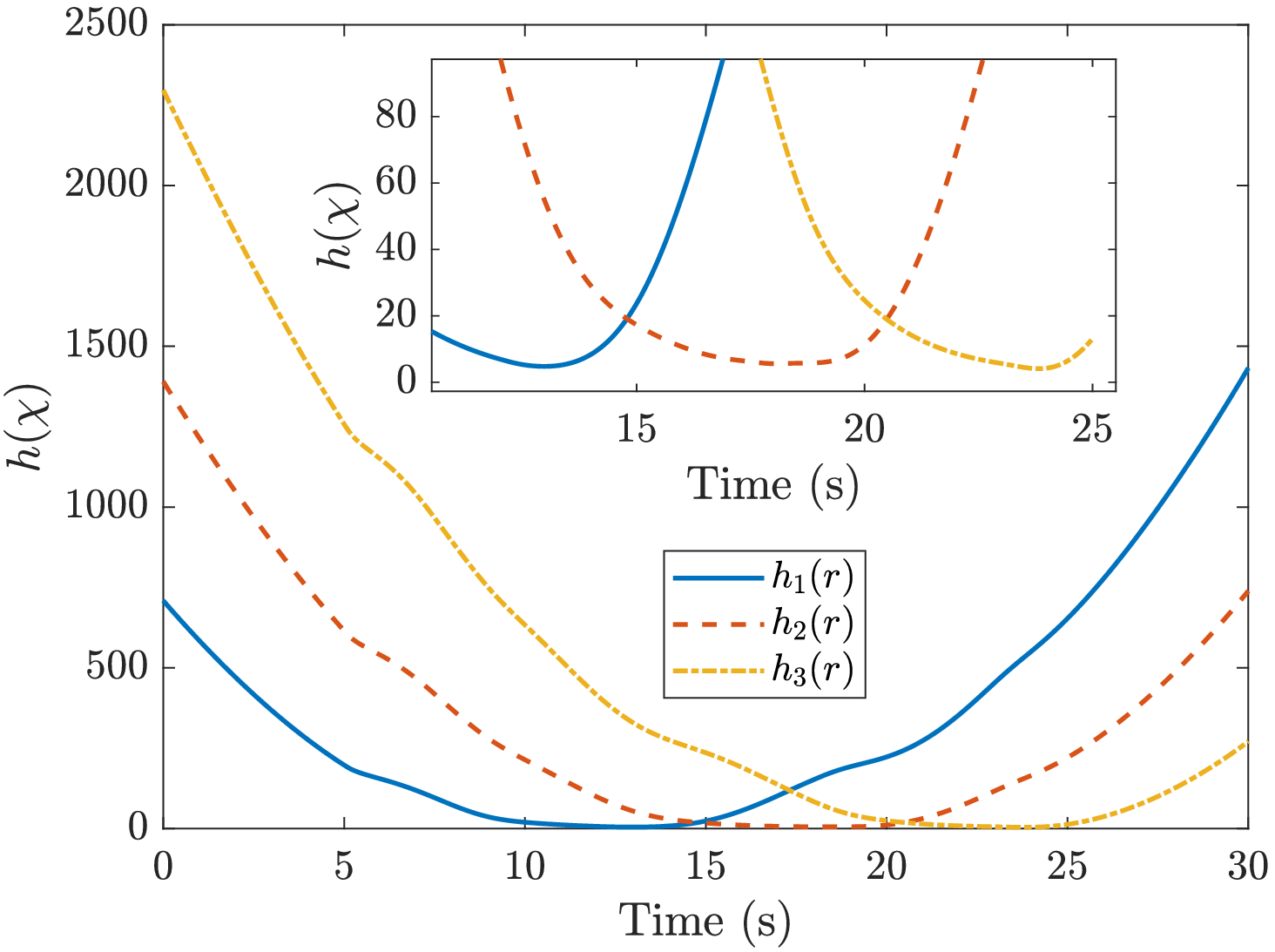}}
	\vspace{-0mm}
	\caption{Results of a straight line tracking task with multiple obstacles. (a) Truck trajectory, (b) Roll angle, (c) CBF value. The horizontal line in (b) denotes $\varphi_{\max}=10$~deg ($\varphi_{\max}+\varphi_G=50$~deg).}
	\label{Fig_Track}
	\vspace{-1mm}
\end{figure*}

We first show the results with different MPC prediction horizons. The situation was setup with an obstacle ($R=2.5$ m and $R_\epsilon =0.5$ m) at location $(5,5)$~m. The vehicle needs to move to $(10, 10)$~m safely by a ski-stunt maneuver. Fig.~\ref{Fig_Diff_H_Traj} shows the trajectory of the vehicle under $H=1,5,10,15$. Fig.~\ref{Fig_Diff_H_Roll} shows the vehicle roll angles, Fig.~\ref{Fig_Diff_H_Steer} shows the controlled steering angles and Fig.~\ref{Fig_Diff_H_h} illustrates the CBF profiles. In all cases, the truck passed the obstacle (Fig.~\ref{Fig_Diff_H_h}) with the CBF applied while closely contacting the buffer zone. Without the buffer zone, the vehicle would collide the obstacle. It is obvious that the vehicle would go through the obstacle area directly if the CBF effect was not applied.

For all successful obstacle avoidance cases, the vehicle trajectories in Fig.~\ref{Fig_Diff_H_Traj} look similar. The differences in roll angle profile are significant as shown in Fig.~\ref{Fig_Diff_H_Roll}. With increased prediction horizon, the roll angle changes become small. Since the desired roll angle was calculated through the BEM, a large roll angle indicates that the curvature of the trajectory is large and therefore it is difficult to follow (large steering angle change, see Fig.~\ref{Fig_Diff_H_Steer}). Table~\ref{Table_Performance} further lists the results that confirm the above analysis. The maximum curvature ($|\rho|_{\max}$) reduces as the prediction horizon increases. The computation cost in each control cycle however increases in this case. For the tradeoff between computation cost and trajectory tracking performance, we chose $H=5$ in the following tests.

\renewcommand{\arraystretch}{1.3}
\setlength{\tabcolsep}{0.1in}
\begin{table}[h!]
	\centering
	\caption{Performance Comparison for Different MPC Horizon}
	\label{Table_Performance}
	\vspace{-1mm}
	\begin{tabular}{|c|c|c|c|c|c|}
		\hline\hline
		$H$& $|\varphi|_{\max}$ & $|\rho|_{\max}$ & $h(\bm \chi)_{\min}$ & Cycle time (ms)   \\ \hline
		$1$ & $8.0$ & $0.23$ & $-0.27$ & $5$ \\	\hline
		$5$ & $7.1$ & $0.21$ & $-0.17$ & $7$ \\	\hline
		$10$ & $5.9$ & $0.17$ & $0$ & $16 $ \\ \hline
		$15$ & $5.1$ & $0.15$ & $0.92$ & $30$ \\ \hline\hline	
	\end{tabular}
	\vspace{-0mm}
\end{table}

We next test the performance of the control strategy at different velocities. An obstacle ($R=2.5$ m and $R_\epsilon =0.5$ m) was set at location $(10, 10)$~m. The vehicle needs to move to $(20, 20)$ m from the origin. Fig.~\ref{Fig_Diff_V} shows the simulation results. With the increased velocity, the vehicle trajectory was further away from the obstacle, as shown in Figs.~\ref{Fig_Diff_V_Traj} and~\ref{Fig_Diff_V_h}. Traveling at a large velocity, the CBF modification effect also becomes significant to prevent any possible collision, comparing with the case at a smaller velocity. One advantage at a large velocity is that the steering effect for balance control is enhanced. From~(\ref{Eq_Steering_Troque}), given the same steering angle, the balance torque increases with large velocity $v$. This explains the results in Fig.~\ref{Fig_Diff_V_Steer} that at $t=0.1$~s the steer angle under $v=5$ m/s is the smallest to balance the vehicle at $-8$~degs.

\setcounter{figure}{3}
\begin{figure}[h!]
	\hspace{-3mm}
	\subfigure[]{
		\label{Fig_Roll_Phase}
		\includegraphics[width=4.3cm]{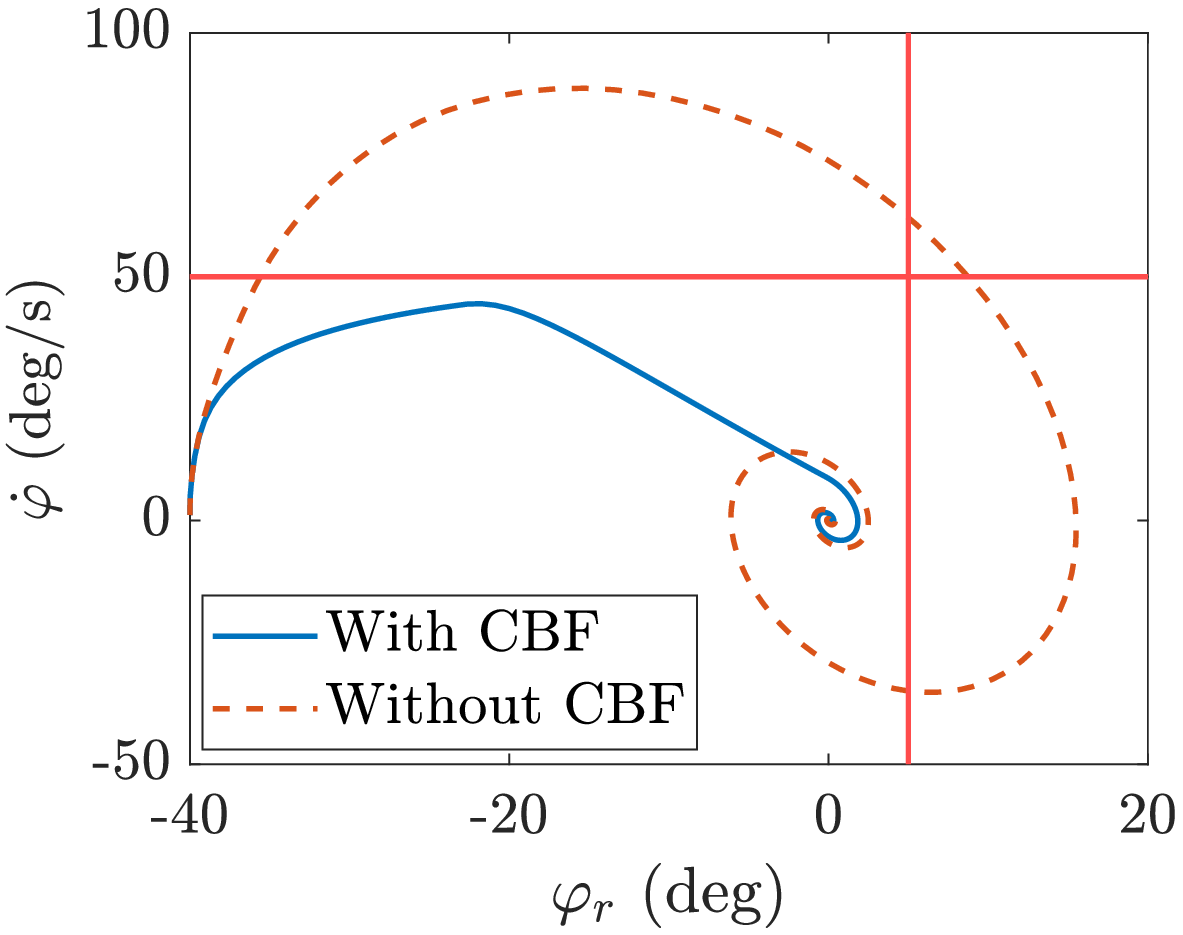}}
	\hspace{-3mm}
	\subfigure[]{
		\label{Fig_Roll_Steer}
		\includegraphics[width=4.3cm]{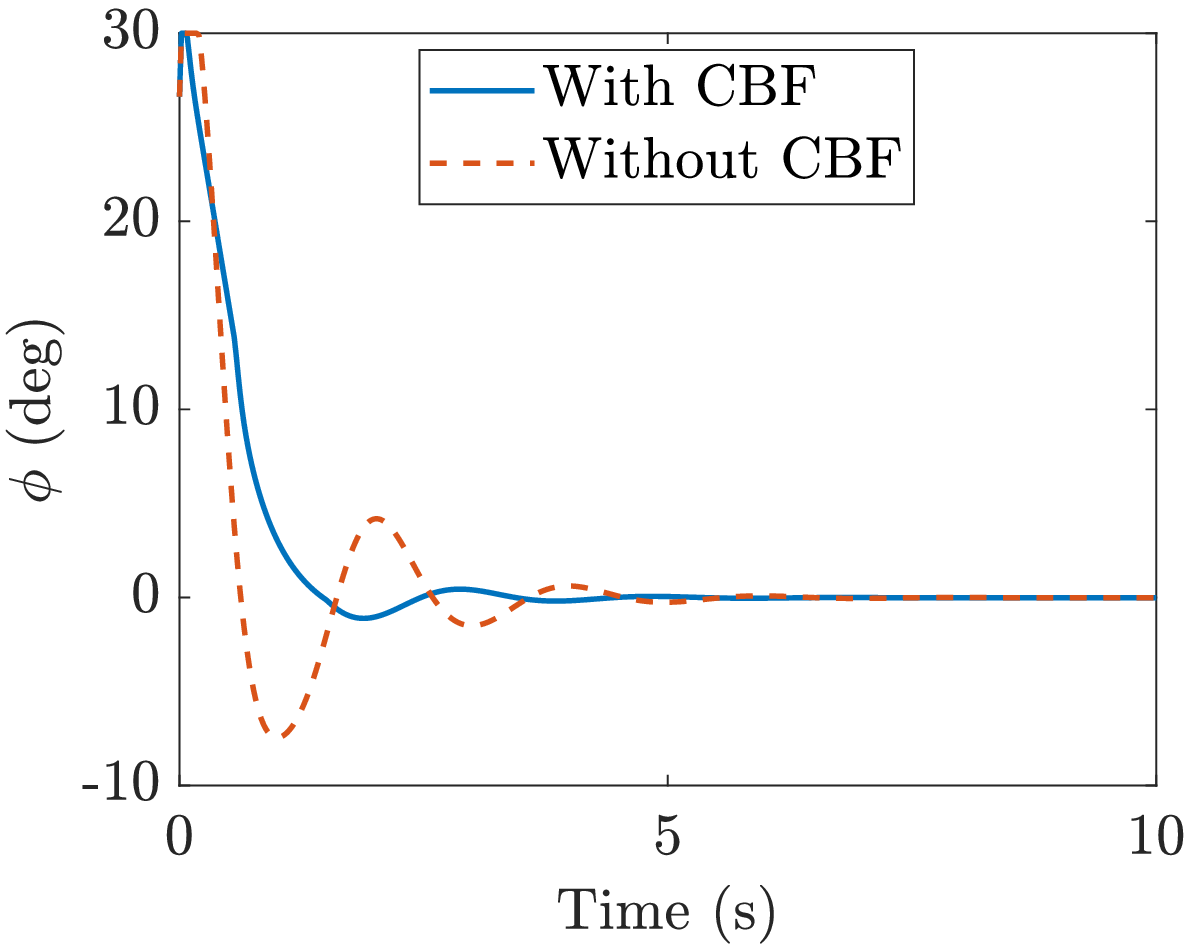}}
	\vspace{-2mm}
	\caption{Initialization of the ski-stunt maneuver. $\varphi_0=-40$~deg ($\varphi_0+\varphi_G=0$~deg) corresponds to the four-wheel driving model. (a) Phase portrait of the roll motion, (b) Steering angle. The horizontal and vertical lines in (a) denote $\dot\varphi_{\max}=50$~deg/s and $\varphi_{\max}=5$~deg ($\varphi_{\max}+\varphi_G=45$~deg), respectively.}
	\label{Fig_Roll}
	\vspace{-3mm}
\end{figure}

\setcounter{figure}{5}
\begin{figure}[ht!]
 \centering
  \includegraphics[width=3.3in]{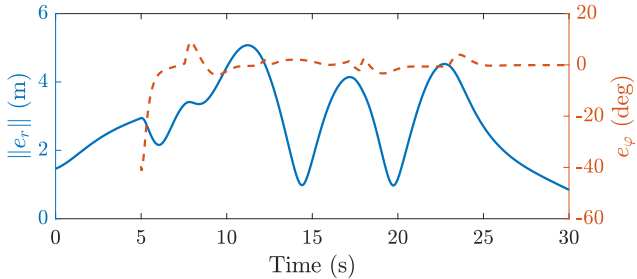}
  \vspace{-0mm}
  \caption{Planar motion and roll motion errors of a straight-line tracking example.}
  \label{Fig_Track_Error}
  \vspace{-0mm}
\end{figure}

\begin{figure*}[th!]
	\hspace{-3mm}
	\subfigure[]{
		\label{Fig_Exp2_Roll}
		\includegraphics[width=4.72cm]{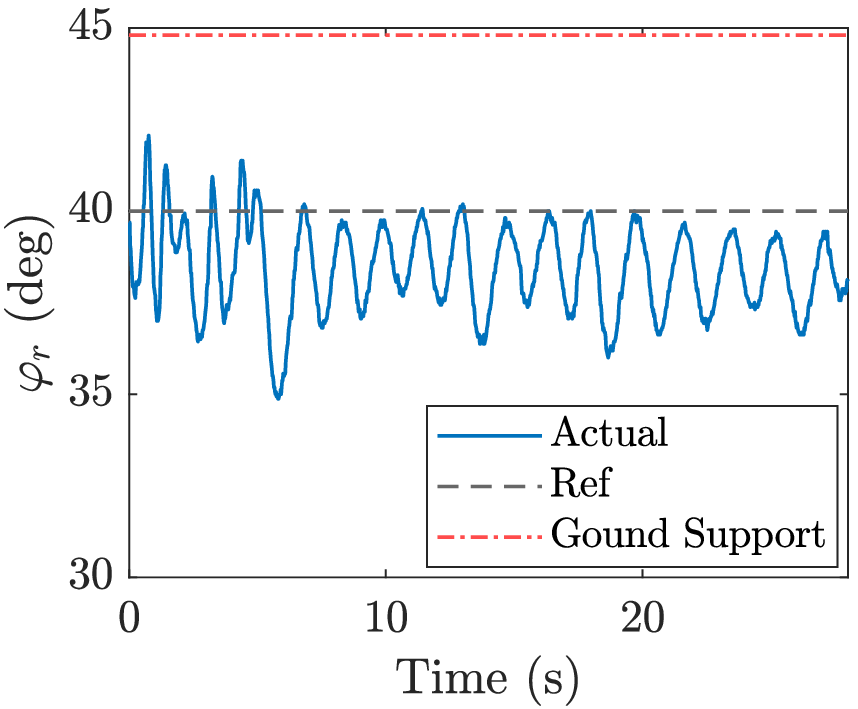}}
	\hspace{-7mm}
	\subfigure[]{
		\label{Fig_Exp2_Steer}
		\includegraphics[width=4.72cm]{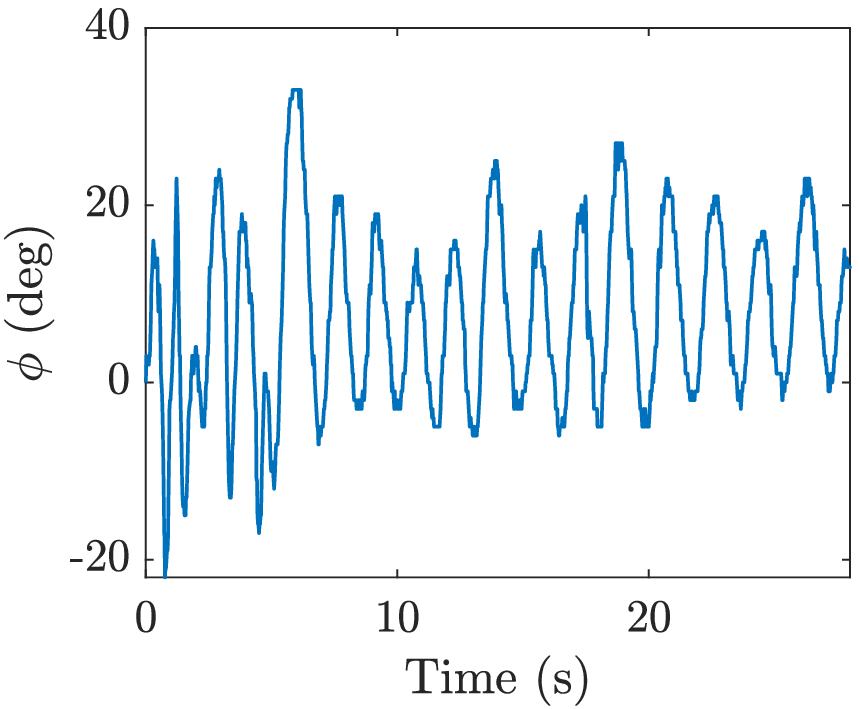}}
	\hspace{-7mm}
	\subfigure[]{
		\label{Fig_Exp2_v}
		\includegraphics[width=4.75cm]{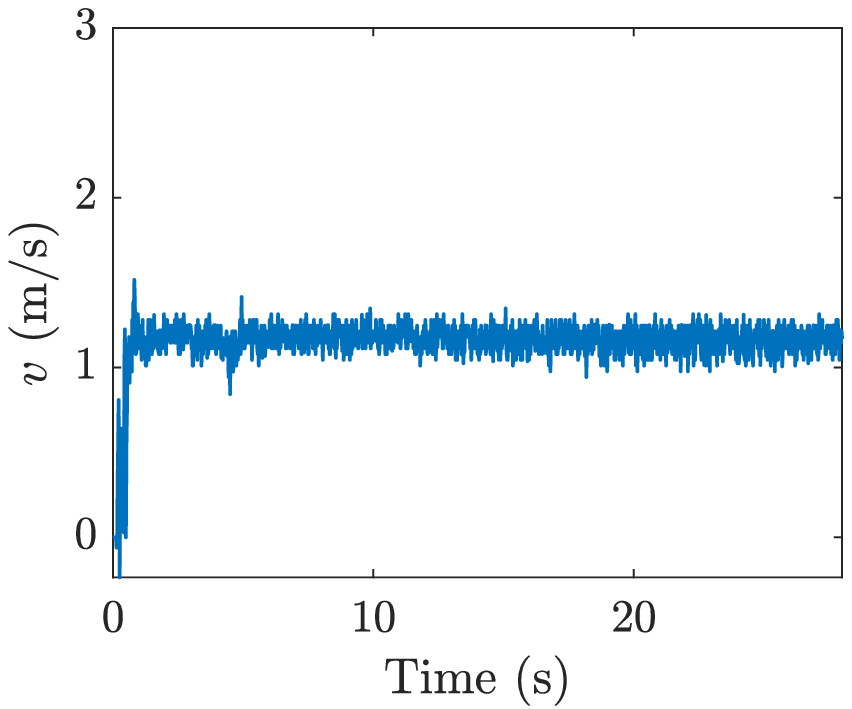}}
	\hspace{-7mm}
	\subfigure[]{
		\label{Fig_Exp2_Traj}
		\includegraphics[width=4.75cm]{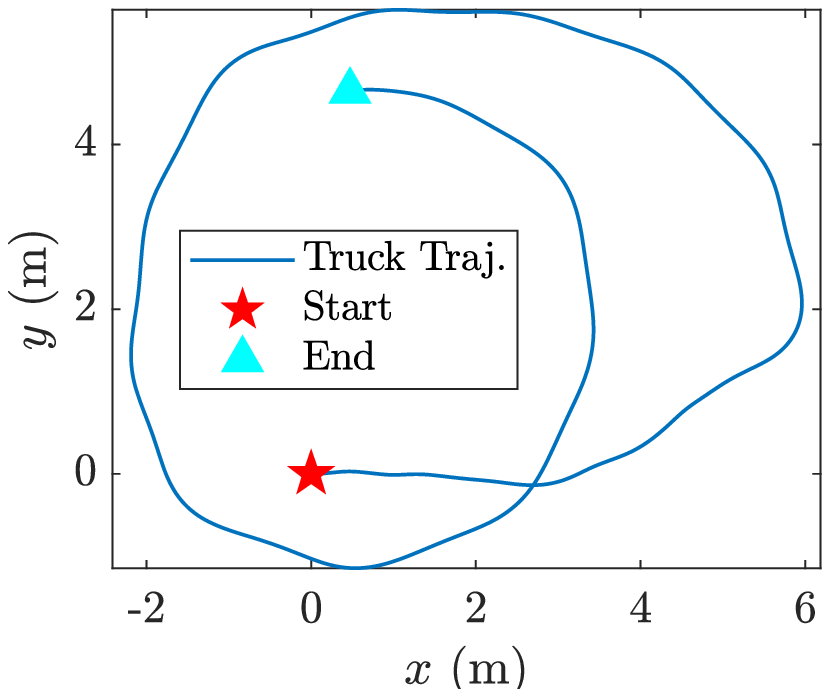}}
	\caption{Balanced ski-stunt maneuver experimental results. (a) Roll angle. (b) Steering angle inputs. (c) Vehicle velocity. (d) Vehicle planar trajectory.}
	\label{Fig_Exp2}
	\vspace{-3mm}
\end{figure*}

Fig.~\ref{Fig_Roll} illustrates the initialization process of the ski-stunt maneuver. The vehicle was at four-wheel driving mode ($\varphi_0=-\varphi_G=-40$~degs) at beginning and the targeted roll angle was around $0$ for the ski-stunt maneuver. The initialization of the ski-stunt maneuvers was created by a sudden steering angle change. To prevent the rollover, $h(\varphi)$ and $h(\dot\varphi)$ were used with $\varphi_{\max}=5$~deg and $\dot\varphi_{\max}=50$~deg/s. The velocity was set at $3$~m/s. Fig.~\ref{Fig_Roll_Phase} shows the roll motion in a phase portrait. With the CBF applied, both the roll angle and roll angular velocity were constrained within the boundaries. For the case without the safe CBF effect, the roll angle reaches $15$~deg and the steer angle displays a large change from $30$ to $-10$~deg. In practice a large positive roll angle in the transient phase indicates a risky situation that the vehicle may roll over completely.

We demonstrate a tracking task with multiple obstacles and Fig.~\ref{Fig_Track} shows the results. The reference trajectory was given by $x_d=1.6t, y_d=1.6t$. Three obstacles were set at at $(20, 20)$, $(27.5, 27.5)$, and $(35, 35)$~m. For safety concern, the maximum roll angle was set at $10$~deg. Thus the CBFs considered were $h_1(\bm r), h_2(\bm r), h_3(\bm r)$ and $h(\varphi)$. Fig.~\ref{Fig_Track_Traj} shows a 3-D illustration with the velocity direction and roll angle direction added to the trajectory profile. From the first $5$~s, the vehicle was in the four-wheel driving mode. At $t=5$~s, the vehicle conducted a sharp turn to initialize the ski-stunt maneuver. Compared with the case without roll motion safety CBF effect, the roll angle (both the BEM $\varphi^e$ and the actual $\varphi$) was less than $10$~deg; see Fig.~\ref{Fig_Track_Roll}. The truck successfully passed three obstacles in an ``$S$''-shape trajectory as shown in Fig.~\ref{Fig_Track_h}. The arrows marked ``Roll'' in Fig.~\ref{Fig_Track_Traj} indicate the roll angle changes to maintain balance. Fig.~\ref{Fig_Track_Roll} shows the reference roll angle (i.e., BEM) and the roll angle closely followed the reference. Fig.~\ref{Fig_Track_Error} shows the planar and roll motion errors and they decay to zero.

\subsection{Preliminary Experimental Result}

As shown in Fig.~\ref{Fig_Truck_Stunt}, a training wheel was added and mounted on one side to protect the vehicle from any damages by possible rollover. When the training wheel touches down on the ground, $\varphi=5$ deg (equivalently $45$~deg rotation from four-wheel driving situation, that is, $\varphi_r=45$ deg). Fig.~\ref{Fig_Exp2} shows the preliminary experiment result to demonstrate of the feasibility of system for a ski-stunt maneuver. The model-based control was used to balance the vehicle and the feedback gains were $k_p=35$ and $k_d=2$.  From the training wheel support stage, the vehicle accelerated and reached to the desired velocity $v=1.2$~m/s. Under the steering input, the vehicle was successfully balanced around $\varphi=-2$~deg (i.e., $\varphi_r=38$~deg) with $\pm 1$ deg oscillation as shown in Fig.~\ref{Fig_Exp2_Roll}. The vehicle trajectory was in a circular shape (radius about $2.5$~m). The results confirmed that with the analytical model and the control design, the vehicle was able to perform the ski-stunt maneuver. Although there exist large modeling errors due to training wheel oscillation, tire deformation, and hardware constraints, etc., the balanced roll angle oscillated and the platform demonstrated feasibility for ski-stunt maneuvers.

\section{Conclusion}
\label{Sec_Conclusion}

This paper studied the aggressive ski-stunt maneuver using a scaled RC truck platform. We considered the trajectory tracking in planar motion and balance of the roll motion under the constraint of underactuated and inherently unstable vehicle dynamics during ski-stunt maneuver. To achieve superior performance, the system model was enhanced by a Gaussian process regression method. We designed a model predictive control that incorporated a probabilistic exponential control barrier function method for collision avoidance and balanced roll motion. Under the proposed control design, the ski-stunt maneuver was proved to be stable and safe. The control algorithm was validated extensively through numerical simulation examples. We also demonstrated the feasibility of the autonomous ski-stunt maneuver using the scaled truck. We are currently working to extend the experiments to demonstrate the performance under the proposed modeling and control design.

\bibliographystyle{IEEEtran}
\bibliography{YiRef}

\end{document}